\documentclass[aps,pre,twocolumn,superscriptaddress,showpacs]{revtex4-1}

\usepackage{amsmath,amssymb,bm}
\usepackage{graphics}
\usepackage{graphicx}
\usepackage{comment}
\usepackage{mathtools}
\usepackage{natbib}

\begin{document}
\title{
Solution of the explosive percolation quest. II. 
Infinite-order transition 
produced by the initial distributions of clusters
}
\author{R.~A.~da~Costa}
\affiliation{Departamento de F{\'\i}sica da Universidade de Aveiro $\&$ I3N, Campus Universit\'ario de Santiago, 3810-193 Aveiro, Portugal}
\author{S.~N. Dorogovtsev}
\affiliation{Departamento de F{\'\i}sica da Universidade de Aveiro $\&$ I3N, Campus Universit\'ario de Santiago, 3810-193 Aveiro, Portugal}
\affiliation{A.F. Ioffe Physico-Technical Institute, 194021 St. Petersburg, Russia}
\author{A.~V. Goltsev}
\affiliation{Departamento de F{\'\i}sica da Universidade de Aveiro $\&$ I3N, Campus Universit\'ario de Santiago, 3810-193 Aveiro, Portugal}
\affiliation{A.F. Ioffe Physico-Technical Institute, 194021 St. Petersburg, Russia}
\author{J.~F.~F. Mendes}
\affiliation{Departamento de F{\'\i}sica da Universidade de Aveiro $\&$ I3N, Campus Universit\'ario de Santiago, 3810-193 Aveiro, Portugal}
\begin{abstract}
We describe the effect of power-law initial distributions of clusters on ordinary percolation and its generalizations, specifically, models of explosive percolation processes based on local optimization. These 
aggregation processes were shown to exhibit continuous phase transitions if the evolution starts from 
a set of disconnected nodes. Since the critical exponents of the order parameter in explosive percolation transitions turned out to be very small, these transitions were first believed to be discontinuous. In this article we analyze the evolution starting from clusters of nodes whose sizes are distributed according to a power law. We show that these initial distributions change dramatically the position and order of the phase transitions in these problems. We find a particular initial power-law distribution producing a peculiar effect on explosive percolation, namely before the emergence of the percolation cluster, the system is in a ``critical phase'' with an infinite generalized susceptibility. 
This critical phase is absent in ordinary percolation models with any power-law initial conditions. 
The transition from the critical phase is an infinite order phase transition, which resembles the scenario of the Berezinskii--Kosterlitz--Thouless phase transition. We obtain the critical singularity of susceptibility at this peculiar infinite-order transition in explosive percolation. It turns out that susceptibility in this situation does not obey the Curie-Weiss law. 
\end{abstract}
\pacs{64.60.ah, 05.40.-a, 64.60.F-}
\maketitle



\section{Introduction}

The percolation transition is the key phase transition occurring in disordered systems including disordered lattices and random networks 
\cite{stauffer1979scaling,stauffer1991introduction,dorogovtsev2008critical,dorogovtsev2010lectures}. 
The gradual increase 
of the number of links in a network or lattice leads to 
the growth of clusters of connected nodes
and eventually to the formation of a percolation cluster (giant connected component) at the percolation threshold. 
This phase transition was studied in detail, and understood to be continuous in all disordered systems which were explored. 
Recently a new class of irreversible percolation processes, so-called ``explosive percolation'', was introduced \cite{achlioptas2009explosive}, where the new links are added to the system using metropolis-like algorithms. 
Although these processes directly generalize ordinary percolation, they demonstrate a set of features remarkably distinct from ordinary percolation. 
The unusual properties of this kind of percolation 
led 
to the initial reports based on simulations 
\cite{achlioptas2009explosive,ziff2010scaling,friedman2009construction,d2010local,ziff2009explosive,cho2009percolation,radicchi2009explosive,radicchi2010explosive,araujo2011tricritical} 
that these processes 
show discontinuous 
transitions. 
Solving the problem analytically, we have shown that the explosive percolation transitions are actually continuous~\cite{da2010explosive}. These transitions have a so small critical exponent of the percolation cluster size, that in simulations of finite systems they can be easily perceived as discontinuous~\cite{da2010explosive,da2014critical}. This conclusion was supported by subsequent works of physicists \cite{nagler2011impact,grassberger2011explosive,lee2011continuity,fortunato2011explosive} and mathematicians \cite{riordan2011explosive}. 

\begin{table*}[t]
\begin{tabular}{c | c | c | c}
\hline
&&&
\\[-10pt]
& 
$\tilde{\tau}<2+1/(2m-1)$ 
& 
$\tilde{\tau}=2+1/(2m-1)$ 
& 
\  $\tilde{\tau}>2+1/(2m-1)$ \ 
\\[3pt]
\hline
  & 
  $t_c=0$  
  & 
  $t_c=0$  
  & 
  $t_c>0$
  \\[3pt]
   &
 $ P(s,0)\sim s^{1-\tilde{\tau}} $
    &
 $ P(s,0)\sim s^{1-\tilde{\tau}} $
    &
 $ P(s,t_c)\sim s^{-3/2} $
  \\[3pt]
   $\phantom{j}  m=1 \phantom{j} $  
& 
$S\sim t^\beta$ 
& 
$S\sim \exp(\text{-const}/t)$  
& 
$S\sim (t-t_c)^\beta$
 \\[3pt]
  & 
  $\beta=(\tilde{\tau}-2)/(3-\tilde{\tau})$  
  & 
  ---  
  & 
  $\beta=1$
  \\[3pt]
   & 
  $\chi\sim t^{-1}$  
  & 
  $\chi\sim t^{-2}$  
  &  
  $\chi\sim |t-t_c|^{-1}$
  \\[5pt]
    \hline
  & 
  $t_c=0$  
  & 
  $t_c>0$  
  & 
  $t_c>0$
  \\[3pt]
     &
 $ P(s,0)\sim s^{1-\tilde{\tau}} $
    &
 $ P(s,t_c)\sim s^{1-\tilde{\tau}} \ln^\lambda s$
    &
 $ P(s,t_c)\sim s^{1-\tau} $
  \\[3pt]
& 
$S\sim t^\beta$ 
& 
\  $S\sim \exp[\text{-const}/(t-t_c)^\mu]$  \ 
& 
$S\sim (t-t_c)^\beta$
 \\[5pt]
 $ m>1$  
  & 
 \   $\beta= \displaystyle{\frac{\tilde{\tau}-2}{1-(2m-1)(\tilde{\tau}-2)}}$  \ 
  &
  $\mu= \displaystyle{\frac{1}{\lambda(2m-1)-1}}$
   & 
\   $\tau-2 \sim \beta\sim e^{-1.43 m}$ \ 
   \cite{da2014solution} \ 
  \\[10pt]
  & 
  $\chi\sim t^{-1}$  
  & 
   \   $\chi \sim (t-t_c)^{-1-\mu}$ \  if \  $t>t_c$  \ 
  &  
  $\chi\sim|t-t_c|^{-1}$
  \\[3pt]
  &
  &
  $\chi=\infty$ \  if \ $t<t_c$ 
  &
\end{tabular}
\caption{Summary of results. The initial distribution $P(s,t=0)\sim s^{1-\tilde{\tau}}$, $\tilde{\tau}>2$, $S$ is the relative size of the percolation cluster, and $\chi$ is the 
susceptibility. At $\tilde{\tau}=2+1/(2m-1)$, we show only the most singular factor of $S$.
}
\label{t1}
\end{table*}

In our previous work \cite{da2014solution} we developed the scaling theory of explosive percolation phase transitions for a wide range of models, explaining the continuous nature of the transitions and their unusual features. We obtained the full set of relevant critical exponents and scaling functions in the typical situation, in which the evolution starts from isolated nodes or clusters with sufficiently rapidly decaying size distribution. 
In our papers \cite{da2014solution,da2010explosive,da2014critical} we employ the following model of explosive percolation, at which the number of nodes $N$ is fixed and links are added one by one. At each step we choose two sets of $m$ random nodes, from each set we select the node that is in the smallest of $m$ clusters, and then we add a new link between these two nodes. 

In the present article we show 
that 
the explosive percolation transition, as well as ordinary percolation, strongly depends on the initial conditions of the process. In particular, 
slowly decaying initial cluster size distributions can change crucially the nature 
of these transitions. 
The effects are interesting and add much to understanding of explosive percolation and other generalizations of ordinary percolation. 
So in the present article we explore in detail the effect of initial conditions on the percolation transitions in systems including ordinary and explosive percolation models. We consider power-law initial cluster size distributions with exponent $\tilde{\tau}$, and for different values of the exponent find a spectrum of distinct critical behaviors. 
Here we introduce the initial cluster size distribution exponent $\tilde{\tau}$ in contrast to the critical cluster size distribution exponent traditionally denoted by $\tau$. 
Because of the power-law critical  
distribution 
we expect that power-law initial conditions produce interesting effects. 
We will indicate the range of $\tilde{\tau}$ where the transition point coincides with the initial moment of the process, $t_c=0$. In particular, for ordinary percolation, if $\tilde{\tau}=3$, the percolation cluster emerges as $S \sim \exp(-\text{const}/t)$, 
where $S$ is the relative size of this cluster. 
In contrast, for explosive percolation, we find that there exists a value of $\tilde{\tau}$ at which the phase transition turns out to be infinite-order and occurs at $t_c>0$. 
In this situation, the system at $t<t_c$ is in the ``critical phase'' with divergent susceptibility. 
We  
also 
find susceptibility at $t>t_c$ for any $m$ and show that 
its critical exponent is nontraditional, differing from the Curie--Weiss law.
The main results of the paper are presented in Table~\ref{t1}.

The paper is organized in the following way. 
Section~\ref{results} outlines our results and methods.  
In Sec.~\ref{ordinary} we consider effect of power-law initial conditions on ordinary percolation ($m=1$), which is the simplest particular case of the more general model analyzed in this work.  
In Sec.~\ref{explosive} we study the effect of initial conditions on the explosive percolation model ($m\geq 2$), which turns out to be principally different from the case of $m=1$.


\section{Results}
\label{results}

To help the reader let us outline the main results of this article. We use the following set of models. 
At each time step a new link connecting two nodes is
added to the network of $N$ nodes.  At each step sample two
times:
(i) choose $m \geq 1$ nodes uniformly at random and compare
the clusters to which these nodes belong; select the
node within the smallest of these clusters;
(ii) similarly choose the second sampling of $m$ nodes and,
again, as in (i), select the node belonging to the smallest
of the $m$ clusters;
(iii) add a link between the two selected nodes thus merging
the two smallest clusters. 
The resulting process is described by the time dependent probability $P(s,t)$ that a randomly chosen node belongs to a finite cluster of size $s$, where the time $t=L/N$, where $L$ is the number of added links (number of steps of the process). We assume that $N$ is infinite. Then this aggregation process is described by the evolution equation 
\begin{equation}
\frac{\partial P(s,t)}{\partial t}
= s \sum_{u=1}^{s-1} Q(u,t)Q(s{-}u,t) - 2 sQ(s,t)
, 
\label{e0010}
\end{equation}
where $Q(s,t)$ is the probability that a cluster chosen to merge is of size $s$. This probability is expressed in terms of $P(s,t)$. In particular, when $m=1$, the distribution $Q(s,t)$ coincides with $P(s,t)$, and the model is reduced to ordinary percolation. 

We focus on the effect of slowly decaying initial distributions $P(s,t=0)$, namely power laws $P(s,0) \cong a_0 s^{1-\tilde{\tau}}$, which can produce $t_c=0$ or, in the case of explosive percolation, a critical phase. 
Using the evolution equation~(\ref{e0010}) we analyze the Taylor expansion $P(s,t)=P(s,0)+A_1(s)t+A_2 t^2+...$ and obtain the scaling form of the distribution $P(s,t) \cong s^{1-\tilde{\tau}} f(st^{1/\sigma})$. 
We obtain the relation between the scaling function $f(x)$ and the size of the percolation cluster $S\sim t^\beta$. 
We demonstrate that when the exponent $\beta$ is integer, there are two contributions to $S$. The first is  the well known contribution determined by the scaling part of $P(s,t)$ (see, for example, the book~\cite{stauffer1991introduction}). However, there is a second, analytic, contribution that was not considered in Ref.~\cite{stauffer1991introduction}. In particular, the exponent $\beta$ is $1$ for ordinary percolation with rapidly decaying $P(s,0)$. In this standard case, it is the combination of these two contributions $S(t) \cong \text{const} (|t-t_c|+t-t_c)$ that produces $S(t<t_c)=0$ and the proper dependence $S(t>t_c)$.

In Table~\ref{t1}, we present a summary of the main results of this article. The first row of the table 
shows the 
critical behaviors of the ordinary percolation model, $m=1$, for different 
$\tilde{\tau}>2$. The values $\tilde{\tau}>3$ result in a percolation phase transition at $t_c>0$ with standard critical exponents. On the other hand, at $2<\tilde{\tau}\leq 3$ the percolation cluster emerges at $t=0$ with critical exponents given in terms of $\tilde{\tau}$. In the marginal case of $\tilde{\tau}=3$ we observe an infinite-order phase transition with a singularity $S\sim \exp(-\text{const}/ t)$. 
This set of the critical singularities of $S$ for $m=1$ 
agree 
with those obtained in  Ref.~\cite{cho2010cluster}. 
 Interestingly, the susceptibility $\chi = \langle s \rangle_P \equiv \sum_s sP(s)$ obeys the Curie-Weiss law $\chi \sim (t-t_c)^{-1}$ in all the considered situations except at $\tilde{\tau}=3$, when  $\chi \sim t^{-2}$ (note that $\tilde{\tau}\leq 3$ gives $t_c=0$). 
 
 The second row of the table presents our results for explosive percolation $m>1$. 
For these $m$, an infinite order phase transition occurs when $\tilde{\tau}=2+1/(2m-1)$ at $t_c>0$. 
Before $t_c$ the system is in the critical phase, in which the susceptibility \cite{da2014solution} diverges, while 
in the percolation phase the critical behavior of susceptibility $\chi(t>t_c)$ differs from the Curie--Weiss law. 
We find that the size distribution of clusters at the point of the infinite-order phase transition is $P(s,t_c)\sim s^{1-\tilde{\tau}} \ln^\lambda s$. We obtain the exponent $\lambda$ close to $1$ solving the evolution equation numerically. 

Finally, we obtain a general relation between the susceptibility and the size of the percolation cluster close to the critical point, 
\begin{equation}
\chi\cong 
\frac{m}{2}\frac{\partial \ln S }{\partial t} 
.
\label{e0020}
\end{equation}
This simple relation is valid for all models and initial conditions considered in this work. 
The detailed derivations of these analytical results are given in subsequent sections.


\section{Effect of initial conditions in ordinary percolation ($m=1$)}
\label{ordinary}

In the case of $m=1$, our process is actually ordinary percolation, that is at each step we chose uniformly at random two nodes and interconnect them. This can be treated as an aggregation process in which at each step two clusters, chosen with probability proportional to their sizes, merge together. This process is described by the probability distribution $P(s,t)$ that a randomly selected node belongs to a cluster of size $s$ at moment $t$ 
(each step increases time $t$ by $1/N$)
. In the infinite system, the evolution of this distribution is described by the following Smoluchowski equation \cite{krapivsky2010kinetic,smoluchowski1916brownsche}: 
\begin{equation}
\frac{\partial P(s,t)}{\partial t}
= s \sum_{u=1}^{s-1} P(u,t)P(s-u,t) - 2 sP(s,t)
\label{e100}
\end{equation}
for a given initial distribution $P(s,t=0)$.
Defining the generating function $\rho(z,t)$ as
\begin{equation}
\rho(z,t)=\sum_{s=1}^\infty P(s,t) z^s
,
\label{e200}
\end{equation}
we can rewrite Eq.~(\ref{e100}) in terms of $\rho$, and analyze the resulting partial differential equation: 
\begin{equation}
\frac{\partial \rho(z,t)}{\partial t}=-2[1-\rho(z,t)]\frac{\partial \rho(z,t)}{\partial \ln z}
\label{e201}
,
\end{equation}
whose solution $\rho(z,t)$ can be obtained from 
\begin{equation}
\ln z =2t(1-\rho)+g(\rho)
,
\label{e300}
\end{equation}
where the function $g(\rho)$ is determined by initial conditions. 

Let us find this function. Our results will be completely determined by the asymptotic of the initial distribution. 
An initial cluster size distribution with a power-law tail, i.e.,  $P(s) \cong a_0 s^{1-\tilde{\tau}}$ for large $s$, leads to the following singularity of the generating function
\begin{eqnarray}
\rho(1)-\rho(z)&&=\sum_{s=1}^{\infty} P(s)(1-z^s)=a_0\sum_s s^{1-\tilde{\tau}}(1-z^{s})
\nonumber
\\[5pt]
&&=a_0(-\ln z)^{\tilde{\tau}-2} \int_0^\infty dx\, x^{1-\tilde{\tau}}(1-e^{-x})
\nonumber
\\[5pt]
&&=-a_0 \Gamma(2-\tilde{\tau})\left(-\ln z\right)^{\tilde{\tau}-2}
,
\label{e400}
\end{eqnarray}
at $z=1$.

We find the function $g(\rho)$, replacing the left-hand side of Eq.~(\ref{e300}) by Eq.~(\ref{e400}) and putting $\rho(1)=1$, since $S=1-\rho(1)=0$ at $t=0$. Then Eq.~(\ref{e300}) becomes
\begin{equation}
\ln z=2t(1-\rho)-\left(-\frac{1-\rho}{a_0\Gamma(2-\tilde{\tau})}\right)^{1/(\tilde{\tau}-2)}
,
\label{e500}
\end{equation}
which is valid when $z\to 1$. Putting $z=1$ and $1-\rho=S$ we get
\begin{equation}
2t S=\left(-\frac{S}{a_0\Gamma(2-\tilde{\tau})}\right)^{1/(\tilde{\tau}-2)}
.
\label{e600}
\end{equation}
The last equation has two solutions, the trivial one $S=0$ and a nontrivial solution
\begin{equation}
S=(2t)^{(\tilde{\tau}-2)/(3-\tilde{\tau})}\left[-a_0\Gamma(2-\tilde{\tau})\right]^{1/(3-\tilde{\tau})}
,
\label{e700}
\end{equation}
If $\tilde{\tau}>3$ then $\Gamma(2-\tilde{\tau})>0$, and Eq.~(\ref{e600}) has only one real solution $S=0$, showing that the transition does not occur at $t_c=0$ for this range of $\tilde{\tau}$. 
In this case, to find a real solution $S>0$ for $t>t_c>0$ 
we must also consider the analytic terms omitted in Eq.~(\ref{e400}). 
For the range $2<\tilde{\tau}<3$, we have $\Gamma(2-\tilde{\tau})<0$ and the solution~(\ref{e700}) is real and positive, that is
 \begin{equation}
S \cong B t^{(\tilde{\tau}-2)/(3-\tilde{\tau})}
,
\label{e800}
\end{equation}
for small $t$, 
and $B=2^{(\tilde{\tau}-2)/(3-\tilde{\tau})}\left[-a_0\Gamma(2-\tilde{\tau})\right]^{1/(3-\tilde{\tau})}$. Therefore, if $\tilde{\tau}<3$, the transition occurs at the initial moment. 
Note that for $\tilde{\tau}<3$, the first moment of the initial distribution diverges. For ordinary percolation this moment has the meaning of susceptibility \cite{stauffer1991introduction}, and its divergence indicated that the poit $t=0$ is indeed the critical point.
We will consider the case of $\tilde{\tau}=3$ separately in Sec.~\ref{2b}.

It is easy to see that the distribution $P(s,t)$ is an analytic function of $t$. (i) The initial distribution $P(s,0)$ has no divergencies at any $s$. (ii) Let us, for a moment assume that $P(s,t)$ at has singularity $t^{\phi}$ with non-integer $\phi$ at $t=0$. Then the lowest non-integer power on the left-hand side of Eq.~(\ref{e100}) is $t^{\phi-1}$, while on the right-hand side the lowest non-integer power is $t^\phi$, which shows that the assumption was not correct. 
Then we can 
write the Taylor expansion of the function $P(s,t)$ around $t=0$:
\begin{equation}
P(s,t)=A_0(s) +A_1(s)t+A_2(s)t^2+A_3(s)t^3+\dots
.
\label{e900}
\end{equation} 
The first term in expansion~(\ref{e900}) is 
the initial distribution, $A_0(s)=P(s,0)$. The coefficient of the second term, $A_1(s)$, is the first derivative $\partial_t P(s,t=0)$, $A_2(s)$ is the second derivative $(1/2)\partial_t^2 P(s,t=0)$, and so on. Then, given an initial distribution $P(s,0)$, we can find the coefficients $A_i(s)$ sequentially differentiating both sides of Eq.~(\ref{e100}). We analyze these coefficients in different ranges of $\tilde{\tau}$.

The remainder of this section is organized in the following way. In Sec.~\ref{2a} we consider the case of $\tilde{\tau}<3$. In this region, we derive the scaling of the distribution $P(s,t)$ containing a scaling function.  
We obtain the relation between the analytical features of this scaling function and the singularity of the relative size $S$ of the percolation cluster at $t=0$. In Sec.~\ref{2b} we consider the case of $\tilde{\tau}=3$. We show that in this situation all the derivatives of $S(t)$ with respect to $t$ are zero at $t=0$ 
and derive the respective scaling function. In Sec.~\ref{2c} we analyze the singularity of the susceptibility of this problem, and its relation with the percolation cluster size $S$.

 
 \subsection{The case of $\tilde{\tau}<3$}
 \label{2a}
 
Let us first consider the case of $\tilde{\tau}< 3$, derive the scaling of the distribution $P(s, t)$, and the critical singularity of the percolation cluster size.  
 
 \subsubsection{Scaling of $P(s,t)$}
 
We consider 
initial configurations without a percolation cluster, so 
$\sum_s A_0(s)=1$.
We find the coefficients $A_n(s)$, $n{>}0$,
using the identity $\partial_t^{(n)} P(s,t=0)\equiv A_n(s)\, n!$ and Eq.~(\ref{e100}). For the second coefficient we have
\begin{equation}
A_1(s)=  s \sum_{u=1}^{s-1}A_0(u)A_0(s-u) -2 sA_0(s)
.
\label{e1100}
\end{equation}
The sum $\sum_{u<s} A_0(u)A_0(s-u)$ cannot be directly 
reduced to the integral $a_0^2 s^{3-2\tilde{\tau}} \int_0^1 dx\, [x(1-x)]^{1-\tilde{\tau}}$ for large $s$, because it diverges at both limits $0$ and $1$. 
So, following our work~\cite{da2014solution}, we rewrite Eq.~(\ref{e1100}) as 
\begin{align}
A_1(s) =&  s \sum_{u=1}^{s-1}\left[A_0(u){-}A_0(s)\right]\left[A_0(s{-}u){-}A_0(s)\right] 
\nonumber
\\[5pt]
& - s (s{-}1) [A_0(s)]^2 + 2 A_0(s)\sum_{u=1}^{s-1} A_0(u) -2  sA_0(s)
\nonumber
\\[5pt]
=&  s \sum_{u=1}^{s-1}\left[A_0(u){-}A_0(s)\right]\left[A_0(s{-}u){-}A_0(s)\right] 
\nonumber
\\[5pt]
& - s (s{-}1) [A_0(s)]^2 - 2 s A_0(s)\sum_{u=s}^{\infty} A_0(u)
,
\label{e1200}
\end{align}
where we took into account the normalization condition $\sum_s A_0(s)=1$. For large $s$, the sums in last equation can be already reduced to integrals, which converge at both limits,   
and $A_0(s)$ can be 
replaced by $a_0s^{1-\tilde{\tau}}$.  
Then we get
\begin{align}
 A_1(s)
&\cong a_0^2 s^{4-2\tilde{\tau}}  \int_{0}^{1} dx\, \left[x^{1-\tilde{\tau}} -1\right] \left[  (1-x)^{1-\tilde{\tau}} -1\right]
\nonumber
\\[5pt]
& - a_0^2  s^{4-2\tilde{\tau}}
-2 a_0^2 s^{4-2\tilde{\tau}}  \int_{1}^{\infty}dx\, x^{1-\tilde{\tau}}
\nonumber
\\[5pt]
&= a_0^{2} \frac{\Gamma(2-\tilde{\tau})^2}{\Gamma(4-2\tilde{\tau})} s^{4-2\tilde{\tau}} 
\equiv a_1 s^{4-2\tilde{\tau}} 
,
\label{e1300}
\end{align}
where we have introduced the coefficient $a_1$.

The case of $\tilde{\tau}=5/2$ is special. 
For this $\tilde{\tau}$ the coefficient 
$a_1$ in Eq.~(\ref{e1300}) is zero, and $A_1$ decays faster than $s^{4-2\tilde{\tau}}$.  To find the asymptotics of $A_1$ in this situation, we must take into account the higher-order terms that were neglected when passing from the sums 
in Eq.~(\ref{e1200}) to the integrals of Eq.~(\ref{e1300}). 
In general, we estimate the difference of the respective integral and sum with arbitrary exponents $\psi$ and $\phi$ 
\begin{align}
&s\int_{0}^{s} du\, [u^{-\psi}{-}s^{-\psi}][(s{-}u)^{-\phi}{-}s^{-\phi}]
\nonumber
\\[5pt]
&-s\sum_{u=1}^{s-1} [u^{-\psi}{-}s^{-\psi}][(s{-}u)^{-\phi}{-}s^{-\phi}]=
O(s^{-\text{min}(\psi,\phi)})
,  
\label{e1490}
\end{align}
where we assume that $0<\psi,\phi<2$. 
In particular, for $A_1$, we have $\psi=\phi=\tilde{\tau}-1$. 
Then, when $\tilde{\tau}=5/2$, 
the large $s$ asymptotics of 
$A_1$  is 
\begin{equation}
A_1(s) \propto  
-s^{1-\tilde{\tau}} \propto -s^{-3/2} \ \mbox{ if } \  \tilde{\tau}=5/2
.
\label{e1500}
\end{equation}

In Appendix~\ref{a400} we 
calculate the asymptotics of $A_2(s)$: 
 \begin{equation}
A_2(s)
\cong a_0^3 \frac{2 \Gamma(2{-}\tilde{\tau})^3}{3 \Gamma(6{-}3\tilde{\tau})}
s^{7-3\tilde{\tau}} 
\equiv a_2 s^{7-3\tilde{\tau}} 
. 
\label{e2100}
\end{equation}
For the particular values $\tilde{\tau}=7/3$ and $8/3$ the coefficient 
$a_2=0$. 
To find the asymptotic behavior of $A_2$ in these cases
it is necessary to consider higher-order terms that were neglected when passing from Eq.~(\ref{e1800}) to Eq.~(\ref{e1900}), similarly to 
how we treated $A_1$. This analysis gives
\begin{equation}
A_2(s) \propto 
\begin{cases} 
-s^{2-\tilde{\tau}} \propto -s^{-1/3} &\mbox{if }\  \tilde{\tau}=7/3,
\\[5pt]
-s^{4-2\tilde{\tau}}  \propto -s^{-4/3} & \mbox{if }\   \tilde{\tau}=8/3
.
\end{cases} 
\label{e2300}
\end{equation}

These calculations can be repeated for the next terms of expansion~(\ref{e900}), and in general we find that the $n$-th term
\begin{equation}
A_n(s)\cong a_n s^{1-\tilde{\tau}+n(3-\tilde{\tau})}
\label{e2400}
\end{equation}
for large $s$. In Appendix~\ref{a300}, we obtain the general expression
\begin{equation}
a_n= \frac{\left[2 a_0 \Gamma(2{-}\tilde{\tau})\right]^{n+1}}{2 \Gamma[(n{+}1)(2{-}\tilde{\tau})](n{+}1)!} 
.
\label{e3720}
\end{equation}
Notice that this expression generalizes the results for $a_1$ and $a_2$, Eqs.~(\ref{e1300}) and~(\ref{e2100}), respectively. 
The coefficients $a_n$ are expressed in terms of $a_0$ and $\tilde{\tau}$ 
and become zero for $\tilde{\tau}=3-1/(n{+}1)$.
For this $\tilde{\tau}$, the coefficient $A_n(s)$ decays as $s^{-1-1/(n+1)}$.

Equation (\ref{e2400}) enables us to write the Taylor expansion of $P(s,t)$ in the form:
\begin{equation}
P(s,t) \cong s^{1-\tilde{\tau}} \sum_{n=0}^{\infty} a_n \left(s^{3-\tilde{\tau}}t\right)^{n} 
= s^{1-\tilde{\tau}} f(st^{1/(3-\tilde{\tau})})
.
\label{e2500}
\end{equation}
The function $f(x)$ is the scaling function of the problem. 
This function
is only analytic at zero for $\tilde{\tau}=5/2$ when coefficients $a_n$ are zero for odd $n$. 
In general, for \mbox{$2<\tilde{\tau}<3$}, 
$f(x)$ is represented as the series 
\begin{equation}
f(x)=a_0+a_1x^{\sigma}+a_2x^{2\sigma}+a_3x^{3\sigma}...
,
\label{e2600}
\end{equation}
where 
\begin{equation}
\sigma=3-\tilde{\tau}
.
\label{e2601}
\end{equation}
Let us estimate the radius of convergence $x_{rc}$ of this series. 
Expansion~(\ref{e900}) is a convergent series for $t<r$, 
\begin{equation}
r=\lim_{n\to\infty} \left|\frac{A_n(s)}{A_{n+1}(s)}\right| \sim \left|\frac{s^{1-\tilde{\tau}+n(3-\tilde{\tau})}}{s^{1-\tilde{\tau}+(n+1)(3-\tilde{\tau})}}\right| \sim s^{\tilde{\tau}-3}
.
\label{e2700}
\end{equation}
Then series~(\ref{e2600}) is convergent for $st^{1/(3-\tilde{\tau})}=x<x_{rc}=sr^{1/(3-\tilde{\tau})}\sim 1$.

\begin{figure}[]
\begin{center}
\includegraphics[scale=0.45]{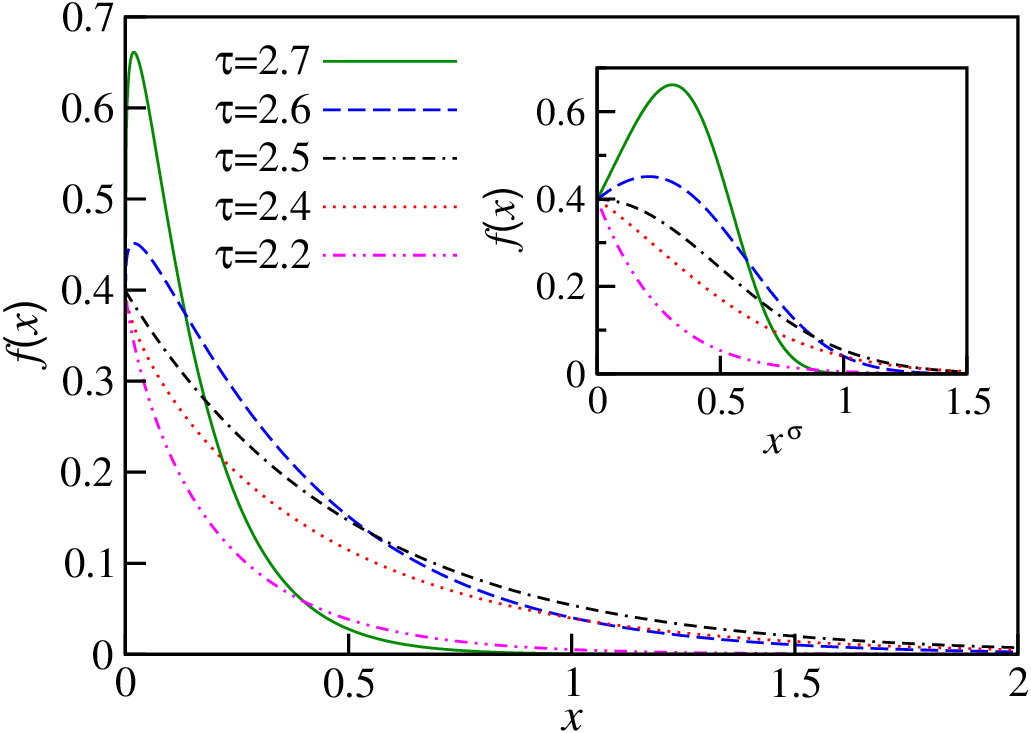}
\end{center}
\caption{Scaling function $f(x)$ calculated using the first $1000$ terms of the series 
(\ref{e2600}) 
for $a_0=1/\sqrt{2 \pi}$ and different values of $\tilde{\tau}$. 
In inset, $f(x)$ is shown 
as a function of 
$x^{\sigma}\equiv s^{3-\tilde{\tau}}t$, where $\sigma=3-\tilde{\tau}$. In this representation the function $f$ is analytic at the origin, 
where 
its derivative at the origin changes signs when $\tilde{\tau}$ crosses $5/2$. 
}
\label{f1}       
\end{figure}

Let us consider, for example, the particular case $\tilde{\tau}=5/2$.
In this case the scaling function, 
after substituting Eq.~(\ref{e3720}) into 
Eq.~(\ref{e2600}), reproduces the known result for the mean-field percolation transition at $t_c>0$~\cite{stauffer1979scaling}:
\begin{equation}
f(x)= a_0 \sum_{n=0}^{\infty} \frac{(-4\pi a_0^2  x)^n}{n!}=a_0 e^{-4\pi a_0^2 x}
.
\label{e3740}
\end{equation}
Note that when $\tilde{\tau}=5/2$ the coefficients $a_n$ in series of Eq.~(\ref{e2600}) with odd $n$ are zero. 
If we set $a_0$ to $1/\sqrt{2\pi}$ in Eq.~(\ref{e3740}), we arrive at the known form 
 $f(x)=\exp(-2x)/\sqrt{2\pi}$ for the 
 percolation process starting from isolated nodes \cite{stauffer1979scaling}. 
Figure~\ref{f1} shows scaling functions for several values of $\tilde{\tau}$. 


\subsubsection{Relation between the singularity of $S$ and the scaling function $f(x)$}

The relative size of the percolation cluster $S$ near $t_c$ is independent of the details of distribution $P(s,t_c)$ in the region of small $s$ (non-scaling region). 
This enables us 
to use the scaling form of distribution $P(s,t)$, Eq.~(\ref{e2500}), to recover expression (\ref{e800})  
for $S(t)$ near $t_c$.  
To this end we start from the definition of $S$
\begin{align}
S&\equiv 1-\sum_s P(s,t),
\nonumber
\\[5pt]
&= 1-\sum_s \left(A_0(s)+A_1(s)t + A_2(s)t^2+...\right)
.
\label{e2800}
\end{align}
At $t=0$ the relative size of the percolation cluster $S=0$, which implies $\sum_s A_0(s)=1$, and so 
we can write 
\begin{equation}
S
=-\sum_{n\geq1} t^n B_n
,
\label{e2900}
\end{equation}
which is the Taylor expansion of $S$ at zero with the coefficients
\begin{equation}
 B_n \equiv\sum_s A_n(s)
 .
 \label{e2901}
 \end{equation}
In Eq.~(\ref{e2900}), the coefficients $B_n \propto \sum_s s^{1-\tilde{\tau}+n(3-\tilde{\tau})}$. 
They diverge if $\tilde{\tau}<3-1/(n+1)$ and converge if $\tilde{\tau} \geq 3-1/(n+1)$. 
(Recall that at $\tilde{\tau} = 3{-}1/(n{+}1)$ the coefficient $a_n$ becomes zero and asymptotics of $A_n(s)$ decays as $s^{-1-1/(n+1)}$.)
For any $\tilde{\tau}<3$ the coefficients $B_{n}$ are infinite for $n>\beta$, where 
\begin{equation}
\beta=\frac{\tilde{\tau}-2}{3-\tilde{\tau}}
.
\label{e3000}
\end{equation} 
The divergence of these coefficients in the Taylor series~(\ref{e2900}) indicates the singularity of $S$ at $t=0$. We extract this singularity 
from series~(\ref{e2900}) in the following way. 
We divide the sum $\sum_{n\geq1} B_n t^n$ into two parts, 
$\sum_{n\leq  n^* } B_n t^n+\sum_{n> n^* } B_n t^n$. 
Here 
\begin{equation}
n^*=\lfloor \beta \rfloor
\end{equation}
denotes the largest integer smaller or equal to $\beta$. 
In the second part we 
replace $A_n$ by their asymptotic form, namely  
\begin{align}
S=&-\sum_{1\leq n\leq  n^* }  B_n t^n-\sum_{n> n^* } \sum_s a_n s^{1-\tilde{\tau}+n\sigma} t^n 
\nonumber
\\[5pt]
&-\sum_{n> n^* } t^n \sum_s \left(A_n(s)- a_n s^{1-\tilde{\tau}+n\sigma}\right)
\nonumber
\\[5pt]
=&-\sum_{1\leq n\leq  n^* }  B_n t^n
- \sum_s  s^{1-\tilde{\tau}}\sum_{n> n^* } a_n (st^{1/\sigma})^{n\sigma} 
\nonumber
\\[5pt]
&-\sum_s s^{-2} \sum_{n> n^* }O[(st^{1/\sigma})^{n\sigma}]
\nonumber
\\[5pt]
\cong &-\sum_{1\leq n\leq  n^* }  B_n t^n
- \sum_s  s^{1-\tilde{\tau}}f^*(st^{1/\sigma})
.
\label{e3100}
\end{align}
Here
we have used the fact, following from Eq.~(\ref{e1490}), that the deviations from the asymptotics $A_n(s)-a_ns^{1-\tilde{\tau}+n(3-\tilde{\tau})}$ are of the order of $s^{1-\tilde{\tau}+(n-1)(3-\tilde{\tau})}$ for large $s$. 
The function $f^*(x)$ 
is a new scaling function obtained from $f(x)$ by subtracting the first $  n^* +1$ terms of its expansion over $x^\sigma$, 
\begin{equation}
 f^*(x)=f(x)-a_0-a_1x^{\sigma}-...-a_{  n^* }x^{ n^*  \sigma}
 , 
 \label{e3200} 
\end{equation} 
$\sigma=3-\tilde{\tau}$. In particular, if $\tilde{\tau}=5/2$, then $\beta=1$, $a_1=0$, and so $f^*(x)=f(x)-a_0$. 
 
In Appendix~\ref{a300} we find $\sum_s A_n(s)\equiv B_n$ employing the generating functions approach:
\begin{equation}
 B_n =
 \begin{cases}
 0 &\mbox{if } n<\beta
 ,
 \\[3pt]
 \displaystyle{-\frac{2^{n} [-a_0 \Gamma(2-\tilde{\tau})]^{n+1}   }{n+1} }
  &\mbox{if } n=\beta
  ,
\\[5pt]
\infty &\mbox{if } n>\beta
, 
\end{cases}
\label{e3900}
\end{equation}
 Therefore the first sum on the right-hand side of Eq.~(\ref{e3100}) is zero, 
except when $\beta$ is integer, equal to $n^*$, and the sum is $B_{n^*}t^{\beta}$. 
Thus, the singularity of the size of the percolation cluster is 
\begin{align}
S&\cong - \sum_s  s^{1-\tilde{\tau}}f^*(st^{1/\sigma}) - B_{n^*} t^{ \beta}
\nonumber
\\[5pt]
&\cong - t^{(\tilde{\tau}-2)/\sigma}\left[ \int_0^{\infty} dx\,  x^{1-\tilde{\tau}}f^*(x)+ B_{n^*}\right]
.
\label{e3800}
\end{align}
Note that the integral $\int_0^{\infty} dx\,  x^{1-\tilde{\tau}}f^*(x)$ converges at the lower limit 
because we have subtracted from $f(x)$ all terms leading to divergence, see Eq.~(\ref{e3200}). 
Recalling that $\sigma=3-\tilde{\tau}$, we arrive at the same singularity $S = B t^{(\tilde{\tau}-2)/(3-\tilde{\tau})}$ as in Eqs.~(\ref{e800}),  
\begin{eqnarray}
&&B=2^{(\tilde{\tau}-2)/(3-\tilde{\tau})} [-a_0\Gamma(2-\tilde{\tau})]^{1/(3-\tilde{\tau})}
\nonumber
\\[5pt]
&&=-\int_{0}^\infty dx\, x^{1-\tilde{\tau}}f^*(x) -  B_{n^*}
.
\label{e4000}
\end{eqnarray}
The first term on the right-hand side of Eq.~(\ref{e3800}) is the singular contribution from the scaling  behavior $P(s,t)\cong s^{1-\tilde{\tau}} f(s t^{1/\sigma})$.
The second term on the right-hand side of Eq.~(\ref{e3800}) is nonzero only when $\beta=(\tilde{\tau}-2)/(3-\tilde{\tau})$ is integer and equal to $n^*$. 
The contribution of this term comes from the finite $s$ region, $- B_{n^*} \sim \sum_s s^{-1-1/(n^*+1)}$, so it is not included in the scaling function $f^*(x)$. This term is an analytic contribution to $S$ at $\tilde{\tau}=3-1/(n+1)$, $n=[1..\infty ]$, including $\tilde{\tau}=5/2$.

Let us consider briefly the ordinary percolation model ($m=1$) with an initial distribution $P(s,t)$ decaying faster than $s^{-2}$. Then the transition takes place at $t_c>0$, and the critical distribution $P(s,t_c)\cong a_0 s^{1-\tau}$ with ${\tau=5/2}$.
In this situation, the Taylor expansion of $P(s,t)$ around $t=t_c$ has coefficients $a_n$ given by Eq.~(\ref{e3720}) with $5/2$ substituted for ${\tilde{\tau}}$. 
As a result, the scaling form of the distribution $P(s,t)$ is
\begin{equation}
P(s,t)\cong a_0 s^{-3/2} e^{-4\pi a_0^2 s (t-t_c)^2}
\label{e3810}
 \end{equation}
 for $t$ approaching $t_c$ from above and below, where the scaling function is $f(x) = a_0 \exp(-4\pi a_0^2 x)$ in both phases. To obtain this result we simply replace $\tilde{\tau}$ by $5/2$ and $t$ by $t-t_c$ in Eqs.~(\ref{e2500}) and (\ref{e3740}). Making the same replacements in Eq.~(\ref{e3800}), we get 
 \begin{eqnarray}
 S
  &\cong& - a_0 \sum_s  s^{-3/2} \left( e^{-4\pi a_0^2 s (t-t_c)^2} {-}1 \right)-  (t{-}t_c) B_1
  \nonumber
\\[5pt]
  &\cong& - |t{-}t_c| a_0 \int_0^{\infty}  dx\, x^{-3/2} \left( e^{-4\pi a_0^2 x} {-}1 \right)-  (t{-}t_c) B_1
  \nonumber
\\[5pt]
  &\cong&   4\pi a_0^2 ( |t-t_c| +  t-t_c )
.
\label{e3820}
\end{eqnarray}
Note that according to Eq.~(\ref{e3900}), for $\tilde{\tau}=5/2$, the coefficient $B_1=\int_0^{\infty}   dx\, x^{-3/2} ( e^{-4\pi a_0^2 x} {-}1 )=-4\pi a_0^2 $.
This equation describes $S$ in both phases, giving $S=0$ for $t<t_c$, and $S=8a_0^2 \pi (t-t_c)$ for $t>t_c$. 

When $\beta=(\tau-2)/\sigma$ is non-integer, 
only the scaling region contributes to the singularity of the percolation cluster size
\begin{equation}
S\cong -|t-t_c|^\beta \int_0^\infty dx\, x^{1-\tau} f^*(x)
.
\nonumber
\end{equation}
In this situation, as is noted in Ref.~\cite{stauffer1991introduction}, the scaling function $f^*(x)$ must be different below and above $t_c$. Namely, in the phase $t>t_c$ the scaling function integral
\begin{equation}
\int_0^\infty dx\, x^{1-\tau} f^*(x)=-B
\nonumber
,
\end{equation}
while
in the phase $t<t_c$ the integral  
\begin{equation}
\int_0^\infty dx\, x^{1-\tau} f^*(x)=0
\nonumber
\end{equation}
to comply with $S=0$. Consequently the 
scaling function $f(x)=f^*(x)+a_0+...+a_{ n^* } x^{ n^*  \sigma}$ must have a maximum $f(x_\text{max}{>}0)>f(0)$ in the phase $t<t_c$~\cite{stauffer1991introduction}.
This asymmetry of the scaling function is observed, for example, for ordinary percolation at dimension $d<6$ \cite{nakanishi1980scaling}, and in explosive percolation \cite{da2010explosive,da2014solution,ziff2010scaling}. In these examples, $0<\beta<1$. 
On the other hand, when $\beta$ is integer, the non-scaling region additionally contributes to the singularity of $S$. This contribution, which is important, in particular, for ordinary percolation above the upper critical dimension $6$ where $\beta=1$, was not considered in the book~\cite{stauffer1991introduction}.  
Thanks to the non-scaling contribution $B_1(t-t_c)$ to $S$, $f(x)$ is the same monotonically decreasing function in both phases in this situation, see Eqs.~(\ref{e3810}) and~(\ref{e3820}).

 
 \subsection{The case of $\tilde{\tau}=3$}
 \label{2b}
 
Now we  show that in the case of $\tau= 3$  all derivatives
of $S(t)$ with respect to $t$ are zero at $t = 0$ and then we find the
respective scaling function. 
In this situation the 
singularity of the generating function $\rho(z,0)$ at $z=1$ differs from that in Eq.~(\ref{e400}), namely 
 \begin{equation}
 1-\rho(z,0)
 \cong -a_0(1-z)\ln(1-z)
.
\label{e4100}
 \end{equation}
 The inverse function is 
 \begin{equation}
\ln z \cong z-1\cong \frac{1-\rho}{a_0 \ln(1-\rho)}
.
\label{e4200}
 \end{equation}
The right-hand side of this expression gives the function $g(\rho)$ in Eq.~(\ref{e300}).  
Using Eqs.~(\ref{e300}) and~(\ref{e4200}) we write the equation for the generating function  $\rho(z,t)$ near 
$z=1$ as follows 
\begin{equation}
\ln z =2t(1-\rho)+ \frac{1-\rho}{a_0 \ln(1-\rho)}
.
\label{e4300}
\end{equation}
At $z=1$ this equation gives the relative size of the percolation cluster for small $t$: 
\begin{equation}
S \sim e^{-1/(2 a_0 t)}
.
\label{e4400}
\end{equation}
Therefore, if $\tilde{\tau}=3$, the transition occurs at the initial moment, and all the derivatives of $S$ are zero at $t=0$, so the percolation transition is infinite-order.


\subsubsection{Scaling of $P(s,t)$}

In the case of $\tilde{\tau}=3$, instead of the Eqs.~(\ref{e2400}) and (\ref{e3720}) for $\tilde{\tau}<3$, 
we derive the general asymptotic expression for the coefficients $A_n(s)$: 
  \begin{equation}
A_n(s) \cong 2^n (n{+}1) a_0^{n+1} s^{-2} \left(\ln s\right)^n
\label{e5100}
\end{equation}
(see Appendix~\ref{a500} for the derivation). 
Within the radius of convergence of the series~(\ref{e900}), this expression gives the following scaling form of the distribution $P(s,t)$:  
 \begin{equation}
P(s,t) 
 \cong s^{-2} f\left( t \ln s \right)
,
\label{e5200}
\end{equation}
where 
\begin{equation}
f(x)=a_0(1-2 a_0 x)^{-2}
.
\label{e5300}
\end{equation}


 \begin{figure}[]
\begin{center}
\includegraphics[scale=0.45]{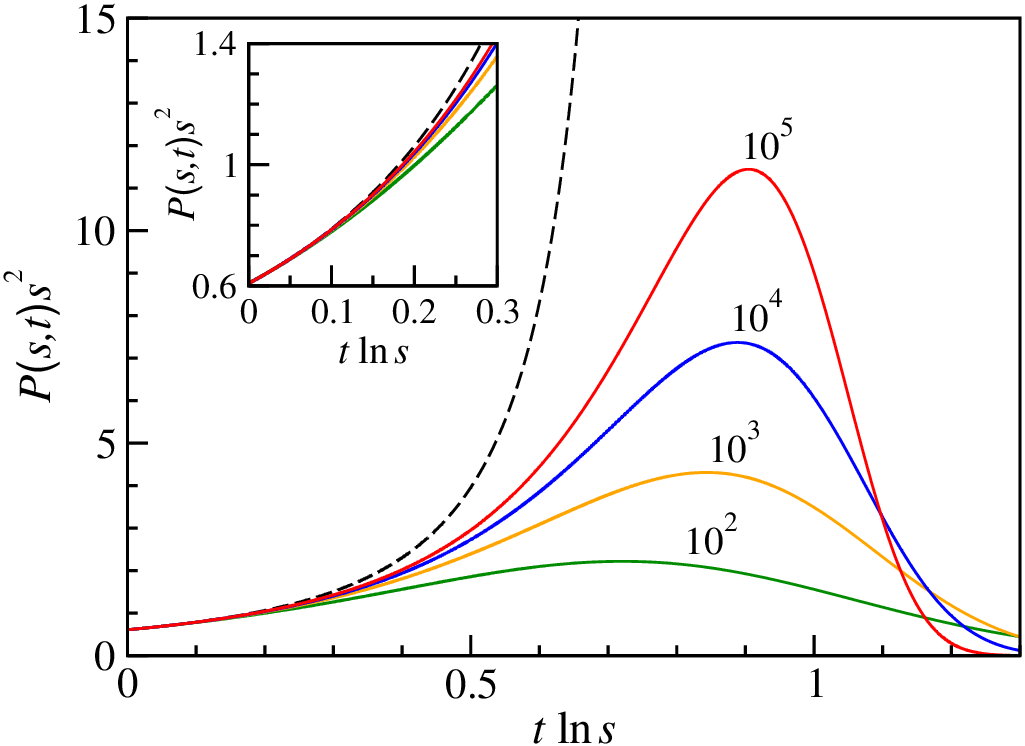}
\end{center}
\caption{
Numerical solution 
of evolution equation~(\ref{e100}) for $s\leq10^5$ with the initial condition $P(s,0)=a_0 s^{-2}$, where $a_0=\zeta(2)^{-1} $. Solid lines, curves $P(s,t)s^2$ vs. $t \ln s$ for the cluster sizes $s$ indicated by the numbers in the plot. Dashed line, scaling function from Eq.~(\ref{e5300}), $f(t \ln s)=a_0/(1-2 a_0 t \ln s)$.
Inset, highlight of the small $t \ln s$ region, in which the curves here depicted approach $f(t \ln s)$.
}
\label{f2}       
\end{figure}


The function $f(x)$ plays the role of a scaling function, which is represented by 
the Taylor series (\ref{e5100}) for $t$ up to the radius of convergence, $t<r=1/(2a_0\ln s)$. Surprisingly, in contrast to the case of $\tilde{\tau}<3$, the scaling function for $\tilde{\tau}=3$ diverges approaching $x_{rc}=1/(2 a_0)$ from below. This divergence requires interpretation, since the distribution $P(s,t)$ itself cannot be divergent at any cluster size, including $s=\exp[(2a_0 t)^{-1}]$. In general, for $t<r$, 
\begin{equation}
\lim_{s\to\infty} P(s,t) s^{\tilde{\tau}-1} \to f(x)
,
\label{e5400}
\end{equation}
where $x$ is a scaling variable such as $|t-t_c| s^{\sigma}$ or $|t-t_c| \ln s$. If the scaling function  converges everywhere in the range of $x$ between $0$ and infinity,  then the curves $P(s,t) s^{\tilde{\tau}-1}$ vs. $x$ will collapse into $f(x)$ at sufficiently large $s$.
 In the 
 case of $\tilde{\tau}=3$ 
the curves $P(s,t) s^{\tilde{\tau}-1}$ vs. $x$ tend to $f(x)$ when $s\to \infty$ in the region $x<x_{rc}$, see Fig.~\ref{f2}. 
These curves have a maximum near 
$x=1/(2 a_0)$. 
As is shown in Fig.~\ref{f2}, the height of this maximum increases, and the difference between $P(s,t) s^{2}$ and $f(t \ln s)$ decreases, as $s$ grows.  
 Note that the curves in Fig.~\ref{f2} tend to $f(t \ln s)$  only for $t \ln s<1/(2 a_0)$, above this point they 
 tend to zero. 
 
In the case of $\tilde{\tau}=3$ we cannot formulate a relation 
similar to Eq.~(\ref{e3800}) between the scaling function $f(x)$ and the singularity of $S(t)$ at $0$ . All the coefficients $B_n$ of the series $S(t)=-\sum_{n\geq1} B_n t^n$  are $0$, which indicates an infinitely smooth singularity of $S(t)$ at 0.


\subsection{The singularity of susceptibility}
\label{2c}

Let us find the singularity of the susceptibility  and its
relation with the percolation cluster size $S$. 
In mean-field models, near the critical point, the susceptibility typically follows the Curie-Weiss law $\chi\sim|t-t_c|^{-1}$. It is easy to see that this is the case for the ordinary percolation model if 
the exponent of the initial cluster size distribution 
$\tilde{\tau} \neq 3$. 
In ordinary percolation the susceptibility is the average cluster size to which a uniformly randomly chosen node belongs \cite{stauffer1991introduction}, namely 
\begin{equation}
\chi=\langle s \rangle_P=\sum_s s P(s)
.
\label{e4500}
\end{equation}
For $m=1$ the singularities of $\chi$ and $S$ can be related in the following way. 
Summing over $s$ both sides of Eq.~(\ref{e100}) gives
\begin{equation}
\frac{\partial S}{\partial t}= 2 S \langle s \rangle_P= 2 S \chi
.
\label{e4600}
\end{equation}
In the case of $\tilde{\tau}<3$ the percolation cluster size $S$ has a power-law singularity with exponent $\beta=(\tilde{\tau}-2)/(3-\tilde{\tau})$, Eq.~(\ref{e800}), then
\begin{equation}
\chi=\frac{1}{2} \frac{\partial \ln S}{\partial t} \cong \frac{\tilde{\tau}-2}{2(3-\tilde{\tau})} t^{-1}
\label{e4700}
\end{equation}
for small $t$. 

At $\tilde{\tau}=3$, we obtain an anomalous singularity of the susceptibility which  
diverges at $t=0$, $\chi(t{=}0) \sim \sum_s s^{-1}$.  Substituting $S \sim e^{-1/(2 a_0 t)}$ [Eq.~(\ref{e4400})] into Eq.~(\ref{e4600}) we find 
\begin{equation}
\chi=\frac{1}{2} \frac{\partial \ln S}{\partial t} \cong \frac{1}{4 a_0} t^{-2}
,
\label{e4800}
\end{equation}
differing from the Curie-Weiss law. 
Note that when $\tilde{\tau}>3$, the initial susceptibility does not diverge, and the transition occurs at  $t_c>0$ with standard exponents. 
When $\tilde{\tau}<3$, the initial susceptibility diverges, 
the transition occurs at $t_c=0$, and the susceptibility demonstrates the standard behavior, $\chi\sim t^{-1}$.


 \section{Effect of initial conditions on explosive percolation ($m>1$)}
 \label{explosive}

In this section we extend the analysis made in the previous section for $m=1$ to $m>1$. 
We consider the following explosive percolation model \cite{da2010explosive,da2014critical,da2014solution}. 
The evolution starts from a given distribution of clusters. At each step, we choose at random two set of $m$ nodes. Then the node in the smallest cluster of each set is selected, and these two nodes are interconnected. The evolution equation for an arbitrary $m$ takes the form:
\begin{equation}
\frac{\partial P(s,t)}{\partial t}
= s \sum_{u=1}^{s-1} Q(u,t)Q(s{-}u,t) - 2 sQ(s,t)
, 
\label{e5500}
\end{equation}
where $Q(s,t)$ is the probability that a chosen node seats in a cluster of size $s$. 
The relation between $Q(s)$ and $P(s)$ is
\begin{align}
Q(s)& = \left[1-\sum_{u=1}^{s-1}P(u)\right]^{m}\!\!-\ \left[1-\sum_{u=1}^{s}P(u)\right]^{m}
\nonumber
\\[5pt]
&\cong m P(s)\left[1-\sum_{u=1}^{s} P(u) \right]^{m-1}
, 
\label{e5600}
 \end{align}
 where the last approximate equality takes place at 
large $s$ \cite{da2014solution}. 
In the phase with the percolation cluster, this relation between distributions $P(s,t)$ and $Q(s,t)$ is simplified:
\begin{equation}
Q(s) \cong mS^{m-1}P(s)
.
\label{e5700}
 \end{equation}
 Then we can write the partial differential equation for the generating function $\rho(z,t)$ defined in Eq.~(\ref{e200}) for any $m$: 
 \begin{equation}
\frac{\partial \rho}{\partial t}=-2  [S(t)]^{2(m-1)} \left[m^2(1{-}\rho) - m(m{-}1) S(t)\right]\frac{\partial \rho}{\partial  \ln z}
.
\label{e5800}
\end{equation}
This partial differential equation can be solved by applying the hodograph transformation \cite{da2014solution}, which leads to the following equation
 \begin{eqnarray}
 &&\!\!\!\!\!
\int_0^t dt'\,\frac{\partial \ln z}{\partial t'} = \ln z 
\nonumber
\\[5pt]
&&\!\!\!\!\!
= 
\int_0^t dt' 2  [S(t')]^{2(m-1)} \left[m^2(1{-}\rho) - m(m{-}1) S(t')\right] 
\nonumber
\\[5pt]
&&\!\!\!\!\!
+ g(\rho)
, 
\label{e5810}
\end{eqnarray}
where the function $g(\rho)$ is determined by initial conditions. 
Similarly to Eq.~(\ref{e500}), we have
 \begin{equation}
g(\rho) = - \left(-\frac{1-\rho}{a_0\Gamma(2-\tilde{\tau})}\right)^{1/(\tilde{\tau}-2)}
.
\label{e5820}
\end{equation}
At $z=1$, the generating function $\rho(1,t)=1-S(t)$, so 
 \begin{align}
&0=2  m^2S(t)\int_{0}^{t} dt' [S(t')]^{2(m-1)} 
\nonumber
\\[5pt]
&- 2 (m^2{-}m) \int_{0}^{t} dt' [S(t')]^{2m-1}-\left[\!{-}\frac{S(t)}{a_0\Gamma(2{-}\tilde{\tau})}\right]^{1/(\tilde{\tau}{-}2)}
. 
\label{e5900}
\end{align}
One can see 
that
\begin{equation}
S=B t^\beta
\label{e5901}
\end{equation}
 is a non-trivial solution of Eq.~(\ref{e5900}) with 
\begin{equation}
\beta=\frac{\tilde{\tau}-2}{1-(2m-1)(\tilde{\tau}-2)}
,
\label{e6000}
\end{equation}
 and 
\begin{equation}
B = \left[  - a_0 \Gamma(2{-}\tilde{\tau})\right]^{\beta/(\tilde{\tau}-2)}\! \left[\!2m\frac{(\tilde{\tau}{-}2)[1{+}(m{-}1)(\tilde{\tau}{-}2)]}{(3-\tilde{\tau})\beta}\!\right]^{\! \beta}
.
\label{e6100}
\end{equation}

In Ref.~\cite{da2014solution} we introduced the generalized susceptibility for these explosive percolation models. The susceptibility $\chi$ is introduced in terms of the probability $c_2$ that two nodes selected by our algorithm belong to the same cluster
\begin{equation}
c_2 = \frac{\chi}{N} + S^{2m} = \frac{1}{N}\sum_s \frac{s Q^2(s)}{P(s)} + S^{2m}
.
\label{e6110}
\end{equation}
In particular, when $m=1$, this susceptibility is reduced to the standard one for ordinary percolation $\chi = \langle s \rangle_P$ \cite{stauffer1991introduction}. 
In \cite{da2014solution} we showed that the susceptibility is divergent at $t=0$ if the exponent of the initial cluster size distribution 
$\tilde{\tau} \leq 2+1/(2m-1)$. For 
$\tilde{\tau}>2+1/(2m-1)$ the transition occurs at $t_c>0$, with the critical exponents and scaling functions calculated in~\cite{da2014solution,da2014critical}. 
If $\tilde{\tau}<2+1/(2m-1)$ the size of the percolation cluster follows the power-law $S\cong B t^\beta$, with $\beta$ and $B$ given by Eqs.~(\ref{e6000}) and~(\ref{e6100}) respectively.
Here the case of $\tilde{\tau}=2+1/(2m-1)$ for $m>1$ distinguishes itself significantly from that for $m=1$. In the next subsection we show that an infinite-order percolation transition takes place at $t_c$ exceeding zero when $\tilde{\tau}=2+1/(2m-1)$ and $m>1$. 

The remainder of this section is organized in the following way. In Sec.~\ref{3a} we consider the region $\tilde{\tau}<2+1/(2m-1)$. For this range of $\tau$ we derive the scaling form of the distribution $P(s,t)$ and relate the scaling function and the singularity of $S$. In Sec.~\ref{3b} we consider the case of $\tilde{\tau}=2+1/(2m-1)$ and show that in this situation we have an infinite-order phase transition. Next we derive scaling functions for this problem and the singularity of the relative size of the percolation cluster $S$. Finally in Sec.~\ref{3c} we find the critical singularity of susceptibility.


 \subsection{The case of $\tilde{\tau}<2+1/(2m-1)$} 
 \label{3a}
 
 Let us first consider the region $\tilde{\tau}<2+1/(2m-1)$, derive the scaling form of the distribution $P(s, t)$, and obtain the critical singularity of the percolation cluster size.
 
  \subsubsection{Scaling of $P(s,t)$}
  
For $m>1$, 
the coefficients $A_n(s)$ of the Taylor expansion of $P(s,t)$, Eq.~(\ref{e900}), 
 can be calculated in a similar way to the case of $m=1$.
 Using Eq.~(\ref{e5500}) at $t=0$ we write 
\begin{eqnarray}
A_1(s)
=&& s \sum_{u=1}^{s-1} Q(u,0)Q(s{-}u,0) - 2 sQ(s,0)
\nonumber
\\[5pt]
=&& s \sum_{u=1}^{s-1} \left[Q(u,0) -Q(s,0)  \right] \left[ Q(s{-}u,0) -  Q(s,0) \right]  
\nonumber
\\[5pt]
&&-s(s-1)Q(s,0)^2-2sQ(s,0)\sum_{u=s}^{\infty} Q(u,0)
,
\label{e6200}
\end{eqnarray}
where we have used the normalization condition $\sum_s Q(s,0)=1$. 
The asymptotics of $Q(s,0)$ are obtained substituting the power-law initial distribution $P(s,0)=A_0(s)\cong a_0 s^{1-\tilde{\tau}}$  into Eq.~(\ref{e5600}):
\begin{equation}
Q(s,0)\cong m A_0(s) \left[\sum_{u>s} A_0(u) \right]^{m-1}  \cong m a_0^m \frac{s^{m(2-\tilde{\tau})-1}}{(\tilde{\tau}{-}2)^{m-1}}
,
\label{e6300}
\end{equation}
Then
the asymptotics of $A_1$ is 
\begin{align}
A_1(s)
\cong &  m^2 a_0^{2m} \frac{s^{2m(2-\tilde{\tau})}}{(\tilde{\tau}{-}2)^{2m-2}} \Bigg[ -1-2 \int_1^{\infty} dx\,  x^{m(2-\tilde{\tau})-1}
\nonumber
\\[5pt]
& + \int_0^{1} dx\, [x^{m(2-\tilde{\tau})-1}{-}1][(1{-}x)^{m(2-\tilde{\tau})-1}{-}1]   \Bigg]
\nonumber
\\[5pt]
=& a_1  s^{2m(2-\tilde{\tau})}
, 
\label{e6400}
\end{align}
where
\begin{equation}
a_1=\frac{a_0^{2m}m^2\Gamma[-m(\tilde{\tau}-2)]^2}{(\tilde{\tau}{-}2)^{2m-2}\Gamma[-2m(\tilde{\tau}-2)]} 
.
\label{e6500}
\end{equation}
For $\tilde{\tau}=2+1/(2m)$ the coefficient $a_1$ becomes zero, and we need to find the next term. 
This term can be obtained by taking into account 
the next-to-leading order term in the expansion of the right-hand side of Eq.~(\ref{e6200}) in powers of $s$.  For this sake we use Eq.~(\ref{e1490}).
In this way, when $a_1=0$, we get the following asymptotics of $A_1(s)\propto s^{m(2-\tilde{\tau})-1}=s^{-3/2}$.

Repeating the procedure for the next coefficients $A_n(s)$ we find that the $n$-th coefficient has the power-law asymptotics: 
\begin{equation}
A_n(s)\cong a_n s^{1-\tilde{\tau}+n\sigma}
,
\label{e6700}
\end{equation}
where the exponent 
\begin{equation}
\sigma=1-(2m-1)(\tilde{\tau}-2)
,
\label{e6800}
\end{equation}
for any $m$. We express the prefactors $a_n$ 
in terms of $m$, $\tilde{\tau}$ and $a_0$, 
similar to $a_1$. For instance, for $a_2$ and $a_3$ we find:
\begin{eqnarray}
a_2=&& \frac{2a_0^{4m-1} m^2 \Gamma[1{-}m(\tilde{\tau}{-}2)]^3 \Gamma[2{-}(3m{-}1)(\tilde{\tau}{-}2)]}{(\tilde{\tau}-2)^{4m-2}  \Gamma[2{-}2m(\tilde{\tau}{-}2)] \Gamma[1{-}(4m{-}1)(\tilde{\tau}{-}2)]}
,\ \ \ \ \ \ 
\label{e6810}
\\[5pt]
a_3=&& \frac{4a_0^{6m-2}m^4 \Gamma[1{-}m(\tilde{\tau}{-}2)]^4\Gamma[3{-}(5m{-}2)(\tilde{\tau}{-}2)]}{3(\tilde{\tau}{-}2)^{6m-4}  \Gamma[2{-}(6m{-}2)(\tilde{\tau}{-}2)]  \Gamma[2{-}2m(\tilde{\tau}{-}2)]^2}
\nonumber
\\[5pt]
&&\times \Bigg[ \Gamma[2{-}(3m-1)(\tilde{\tau}{-}2)] \Bigg(\frac{\Gamma[2{-}(3m-1)(\tilde{\tau}{-}2)]}{\Gamma[3{-}(5m-2)(\tilde{\tau}{-}2)]}
\nonumber
\\[5pt]
&&-\frac{ [1{-}(4m{-}1)(\tilde{\tau}{-}2)] \Gamma[2{-}2m(\tilde{\tau}{-}2)]}{m(\tilde{\tau}-2)\Gamma[3{-}(4m{-}1)(\tilde{\tau}{-}2)]}\Bigg)
\nonumber
\\[5pt]
&&-(m{-}1)(\tilde{\tau}{-}2)\Gamma[{-}m(\tilde{\tau}{-}2)] \Bigg]
.
\label{e6820}
\end{eqnarray}
When $m>1$, we have to derive expressions for the coefficients $a_n$ individually, unlike the general expression (\ref{e3720}) for all $a_n$ in the case of $m=1$. 
Each coefficient $a_{n>0}$ becomes zero when $\tilde{\tau}=2+n/[n(2m{-}1){+}1]$. 
In this case, the resulting asymptotics of $A_n(s)$ decay as $s^{-1-[n(m-1)+1]/[n(2m-1)+1]}$.

The sums $\sum_s A_{n\geq1}(s)\equiv B_n$ in the equations for $A_{n\geq2}(s)$, are obtained by the generating function approach explained in Appendix~\ref{a300} in detail for $m=1$. 
We find the singularity 
\begin{equation}
\rho_n(z)\cong a_n\Gamma[2{-}\tilde{\tau}{+}n\sigma] (1{-}z)^{\tilde{\tau}-2-n\sigma}
,
\label{e6830}
\end{equation}
by differentiating the evolution equation~(\ref{e5500}) and relation~(\ref{e5600}).
The value of $B_n=\rho_n(1)$ diverges or converges depending on $\tilde{\tau}$, similarly to the case of $m=1$:
\begin{equation}
B_n=
\begin{cases}
0 &\mbox{if } n<\beta
\\[5pt]
a_n\Gamma[2{-}\tilde{\tau}{+}n\{1{-}(2m{-}1)(\tilde{\tau}{-}2)\}] &\mbox{if }  n=\beta \ \ \ \ \
\\[5pt]
\infty &\mbox{if } n>\beta
\end{cases}
\label{e6840}
\end{equation}
where
$\beta=(\tilde{\tau}-2)/[1-(2m-1)(\tilde{\tau}-2)]$.

The function $P(s,t)$ can be written in the scaling form:
\begin{equation}
P(s,\delta) \cong s^{1-\tilde{\tau}} \sum_{n=0}^{\infty} a_n \left(s^\sigma \delta\right)^{n} 
= s^{1-\tilde{\tau}} f(s\delta^{1/\sigma})
.
\label{e6900}
\end{equation}
Here we used the asymptotic behavior of the coefficients $A_n(s)$.
In the vicinity of $x=0$, the expansion of the function $f(x)$ is 
\begin{equation}
f(x)=a_0+a_1x^{\sigma}+a_2x^{2\sigma}+a_3x^{3\sigma}...
,
\label{e7000}
\end{equation} 
where $\sigma=1{-}(2m{-}1)(\tilde{\tau}{-}2)$
  and the coefficients $a_n$ are given by Eqs.~(\ref{e6500}), (\ref{e6810}), (\ref{e6820}), etc.
The scaling function and $S$ are interrelated in the same way as in Eq.~(\ref{e3800}) from Sec.~\ref{2a}, 
\begin{equation}
S
\cong - \delta^{(\tilde{\tau}-2)/\sigma}\left[ \int_0^{\infty} dx\,  x^{1-\tilde{\tau}}f^*(x)+ B_{n^*}\right]
,
\label{e7100}
\end{equation}
where $\sigma=1-(2m-1)(\tilde{\tau}-2)$, $\beta=(\tilde{\tau}-2)/\sigma$, and $n^*=\lfloor \beta \rfloor$. The function  $f^*(x)=f(x)-\sum_{n\leq  n^* } a_n x^{n\sigma}$ as in Eq.~(\ref{e3200}). The constant 
\begin{equation*}
B_{n^*}=\sum_s A_{n^*}(s)=\rho_{ n^*}(1)=a_{n^*} \Gamma\left(2-\tilde{\tau}+ n^* \sigma\right)
\end{equation*}
in nonzero only when $\beta$ is integer, $\beta = \lfloor \beta \rfloor \equiv n^* $, see Eq.~(\ref{e6840}).


\subsection{The case of $\tilde{\tau}=2+1/(2m-1)$}
\label{3b}

Let us consider 
the case of $\tilde{\tau}=2+1/(2m-1)$ and show that in this situation we have an infinite-order phase transition.
When $m>1$, the coefficients $A_n$ in the case of $\tilde{\tau}=2+1/(2m-1)$ have exactly the same form (\ref{e6700}) as for $\tilde{\tau} < 2+1/(2m-1)$. This coincidence is due to the convergent convolution integral in Eq.~(\ref{e6400}) for $A_1$ and similar integrals for $A_{n>1}$. 
This is in contrast to $m=1$, $\tilde{\tau} =3$, where we have the coefficients $A_n \sim  s^{-2}\left(\ln s\right)^n$ with logarithmic factors unlike $A_n \sim  s^{1-\tilde{\tau}+n\sigma}$ for $m=1$, $\tilde{\tau} <3$, see Sec.~\ref{2b}. Recall that these logarithms emerged due to the divergent convolution integral in Eq.~(\ref{e1300}) for $A_1$ at $\tilde{\tau} =3$ and similar integrals for $A_{n>1}$. In this respect, there is a principle difference between $m=1$ and $m>1$. 
 Below we show that for $m>1$ and $\tilde{\tau}=2+1/(2m-1)$ an infinite-order percolation transition takes place at a $t_c>0$.

 
 \subsubsection{Expansion of $P(s,t)$ at $t=0$}
 
\begin{figure}[]
\begin{center}
\includegraphics[scale=0.45]{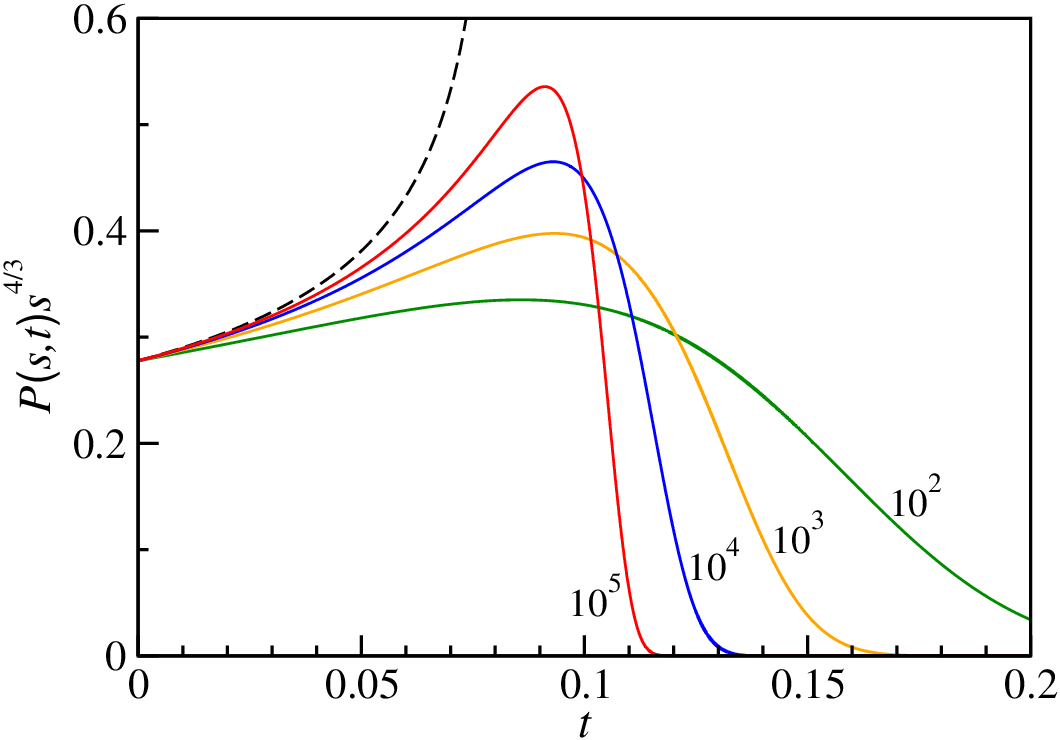}
\end{center}
\caption{
Numerical solution 
of the evolution equation~(\ref{e5500}), for $m=2$ and $s\leq10^5$ with the initial condition $P(s,0)=a_0 s^{-4/3}$, where $a_0=\zeta(4/3)^{-1} $.
Solid lines, curves $P(s,t)s^{4/3}$ vs. $t$ for the cluster sizes $s$ indicated by the numbers in the plot. Dashed line, function $f(t)$ in Eq.~(\ref{e7800}).
}
\label{f3}       
\end{figure}

 When $\tilde{\tau}=2+1/(2m-1)$, we have the exponent $\sigma=0$, so the coefficients $A_n(s)\propto s^{1-\tilde{\tau}}$ for all $n$, see Eq.~(\ref{e6700}). Consequently  
the asymptotics of the cluster size distribution is 
 \begin{equation}
 P(s,t)\cong s^{-2m/(2m-1)} \sum_{n=0}^\infty a_n t^n= s^{-2m/(2m-1)} f(t)
 . 
 \label{e7700}
 \end{equation}
Substituting $\tilde{\tau}-2=1/(2m-1)$ into Eqs.~(\ref{e6500}), (\ref{e6810}), (\ref{e6820}), etc., we find the general expression for the coefficients $a_n$ 
\begin{align}
a_n=&
 \frac{a_0\Gamma[n{+}1/(2m{-}1)]}{n!\,\Gamma[1/(2m{-}1)]}
\nonumber
\\[5pt]
&\times \left( \frac{2m [a_0(2m{-}1)]^{2m-1} \Gamma[(m{-}1)/(2m{-}1)]^{2}}{\Gamma[2(m{-}1)/(2m{-}1)]} \right)^n
,
 \label{e7710}
 \end{align}
and so 
\begin{equation}
f(t)=\sum_{n=0}^{\infty} a_n t^n=
 a_0  \left( 1{-} \frac{t}{r} \right)^{1/(1-2m)}
,
\label{e7800}
\end{equation}
where
\begin{equation}
r=\frac{ (2m{-}1)^{1-2m} \Gamma[2(m{-}1)/(2m{-}1)] } {2 a_0^{2m-1} m  \Gamma[(m{-}1)/(2m{-}1)]^{2}}
.
\label{e7810}
\end{equation} 
The function $f(t)$ diverges
 at $t=r$.
Fig.~\ref{f3} demonstrates how the curves $P(s,t)s^{2m/(2m-1)}$ vs. $t$ approach $f(t)$ in the region $t<r$ as $s$ approaches infinity.


Substituting Eq.~(\ref{e7700}) into Eq.~(\ref{e5600}), we obtain the asymptotics of the distribution $Q(s,t)$: 
\begin{equation*}
Q(s,t)\cong m (2m{-}1)^{m-1} f(t)^m s^{-1-m/(2m-1)}
.
\end{equation*}
The first moment of this distribution  
\begin{equation*}
 \langle s \rangle_Q=\sum_s s Q(s,t)\sim f(t)^m \sum_s s^{-m/(2m-1)}
 \end{equation*}
  is divergent in the interval $0\leq t\leq r$. Note that also  
  $\langle s\rangle_P=f(t)\sum_s s^{-1/(2m-1)} $ and $\chi \sim f(t)^{2m-1} \sum_s s^{-1}$ are divergent in this interval.   
 Summing over $s$ both sides of Eq.~(\ref{e5500}) we get 
\begin{equation}
\frac{\partial S}{\partial t}=2 S^{m} \langle s\rangle_Q
.
\label{e8200}
\end{equation}
Due to the divergence of $\langle s\rangle_Q$, the only solution of this equation in the interval  $t\leq r$ is $S=0$.
Thus we have a phase without percolation and with divergent susceptibility, which enables us to call this phase ``critical''. 
Below we show that the percolation threshold $t_c$ of this transition is 
exactly $r$. 
For $t>r$, we show in Appendix~\ref{a600} that the moments $\langle s \rangle_Q$ and $\langle s \rangle_P$ are finite.


\begin{figure}[]
\begin{center}
\includegraphics[scale=0.45]{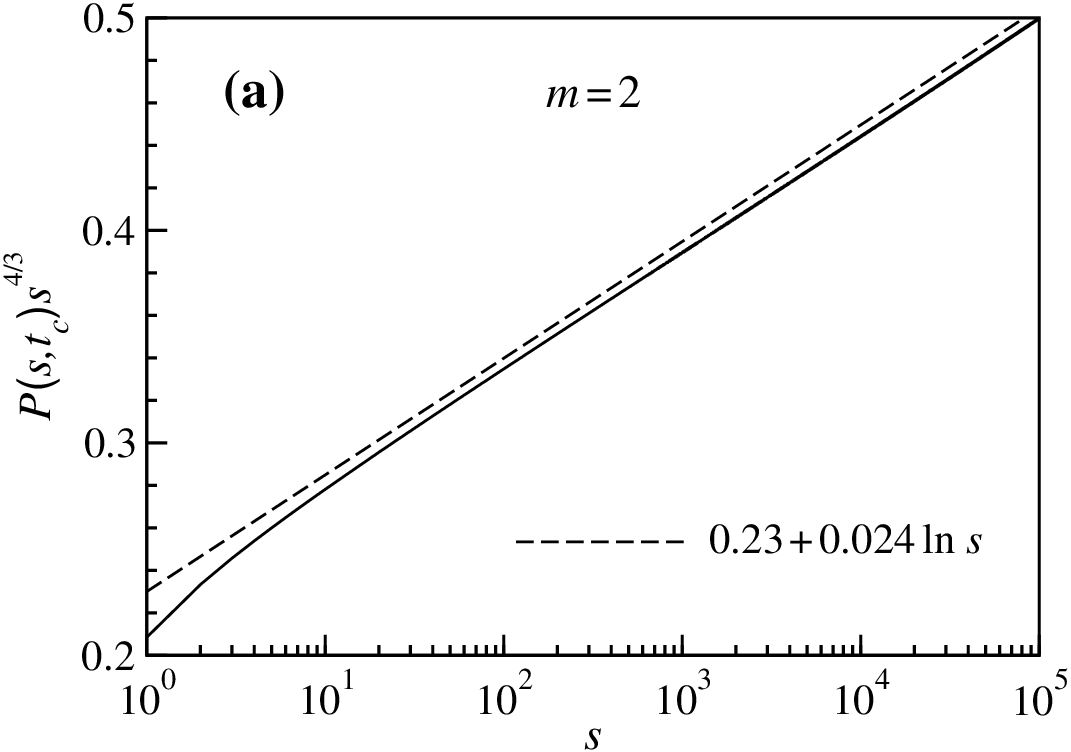}
\\[15pt]
\includegraphics[scale=0.45]{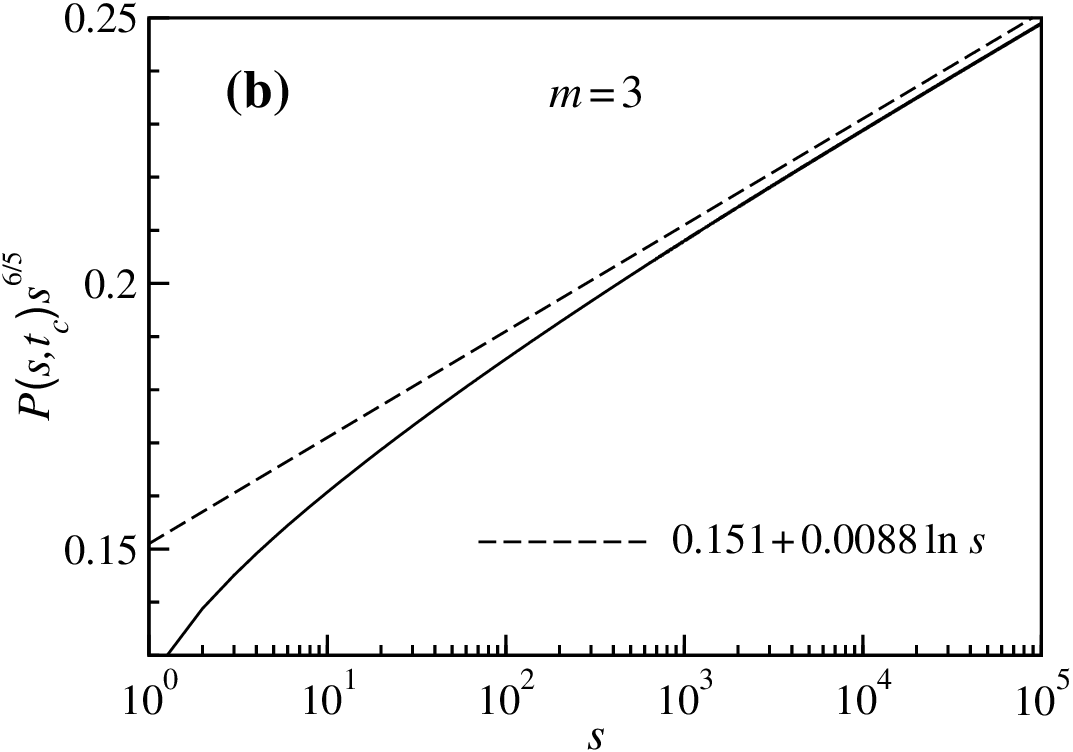}
\end{center}
\caption{
Logarithmic contribution to the asymptotics of $P(s,t_c)$ for (a) $m=2$ and (b) $m=3$. Solid lines, $P(s,t_c)s^{2m/(2m-1)}$ for $s$ from $1$ to $10^5$ 
found numerically from the evolution equations with the initial condition $P(s,0)=a_0 s^{-2m/(2m-1)}$, where $a_0=\zeta[2m/(2m-1)]^{-1}$. Dashed lines, straight lines presented for reference.
}
\label{f4}       
\end{figure}



\subsubsection{Scaling of $P(s,t)$ at $t_c$}

For $t<t_c$ the function $f(t)$ is convergent, and for large 
$s$ the curves $P(s,t)s^{2m/(2m{-}1)}$ collapse into the function $f(t)$, see Fig.~\ref{f3}. 
The value of $P(s,t_c)s^{2m/(2m{-}1)}$ 
approaches infinity as $s\to \infty$. Our numerical results, Fig.~\ref{f4}, indicate that $P(s,t_c)s^{2m/(2m{-}1)}$ grows linearly with $\ln s$ for sufficiently large $s$.
In the range $s\in [1..10^5]$, however, it may be difficult to distinguish different slowly varying functions, such as powers of logarithm. Therefore, we cannot exclude the possibility that the solid line asymptotics follows a law $(\ln s)^\lambda$ with exponent $\lambda$ 
close to but 
different from $1$.

In Appendix~\ref{a100} we obtain the following scaling form of $P(s,t)$ near $t_c$: 
\begin{equation}
P(s,t)\cong C (\ln s)^\lambda s^{-2m/(2m-1)} f\left[C (\ln s)^{\lambda(2m-1)} (t-t_c) \right]
,
\label{e8840}
\end{equation}
where $C$ is a constant, $\lambda\geq 1/(2m{-}1)$, and $f(x)$ is the function defined in  Eq.~(\ref{e7800}).  
Then, the critical distribution $P(s,t=t_c)$ behaves as 
\begin{equation}
P(s,t_c)\cong a_0 C (\ln s)^\lambda s^{-2m/(2m{-}1)}
\label{e8831}
\end{equation}
for large $s$. 
Recall that $P(s,0) \cong a_0 s^{-2m/(2m{-}1)}$.


\subsubsection{Critical behavior of $S$
}

In Appendix~\ref{a200} we derive the critical singularity of the percolation cluster size $S$:  
 \begin{equation}
S \sim \exp(-d \delta^{-\mu})
,
\label{e8833}
\end{equation}
where $\delta = t-t_c>0$, 
\begin{equation}
\mu=\frac{1}{\lambda(2m-1)-1}
,
\label{e8835}
\end{equation}
 and $d$ is a constant
 \begin{equation}
d\! =\!\!  \left(\frac{m(3m{-}2)\!  \left[\lambda{-}1/(2m{-}1)\right]^{1-\lambda(2m-1)}}{(m{-}1)\{-a_0 C \Gamma[-1/(2m{-}1)]\}^{1-2m}}\right)^{1/[1-\lambda(2m-1)]}
.
\label{e8837}
\end{equation}
Note that $\lambda$ cannot be smaller than $1/(2m-1)$.

 
\subsection{The singularity of susceptibility}
\label{3c}

Let us  we find the critical singularity of
susceptibility.  

The critical behavior of $\chi$ is determined by large $s$. 
In the region in which $S>0$, i.e. $t>t_c$, we can substitute $Q(s) \cong mS^{m-1}P(s)$, Eq.~(\ref{e5700}), into Eq.~(\ref{e6110}), which gives 
\begin{equation}
\chi \cong m^2 S^{2m-2}\sum_{s} sP(s,t)=m^2 S^{2m-2}\langle s \rangle_P
\label{e8900}
\end{equation}
close to $t_c$. 
 Summing Eq.~(\ref{e5500}) over $s$ gives
\begin{equation}
\frac{\partial S}{\partial t}=2S^{m}\sum_s s Q(s,t)\cong 2 m S^{2m-1} \langle s \rangle_P
. 
\label{e9000}
\end{equation}
Combining the last two equations we finally get
\begin{equation}
\chi\cong \frac{m}{2 S}\frac{\partial S}{\partial t} =\frac{m}{2}\frac{\partial \ln S }{\partial t} 
.
\label{e9100}
\end{equation}
This remarkably general formula relates the critical singularities of the susceptibility and the percolation cluster size in all situations shown in Table~\ref{t1}. These situations include the finite- and infinite-order continuous phase transitions. 

For $\tilde{\tau}<2+1/(2m{-}1)$, when $S\cong B t^\beta$, this equation ensures that the susceptibility has the following singularity:
\begin{equation}
\chi \cong \frac{m \beta}{2} t^{-1}
.
\label{e9210}
\end{equation}
For $m>1$ and $\tilde{\tau}=2+1/(2m{-}1)$ 
the susceptibility is divergent below $t_c$, while above $t_c$ it has a power-law singularity 
\begin{equation}
\chi \cong \frac{m d \mu}{2} \delta^{-\mu-1}
,
\label{e9300}
\end{equation}
which we obtained by substituting Eq.~(\ref{e8833}) into Eq.~(\ref{e9100}).

\begin{figure}[h]
\begin{center}
\includegraphics[scale=0.4]{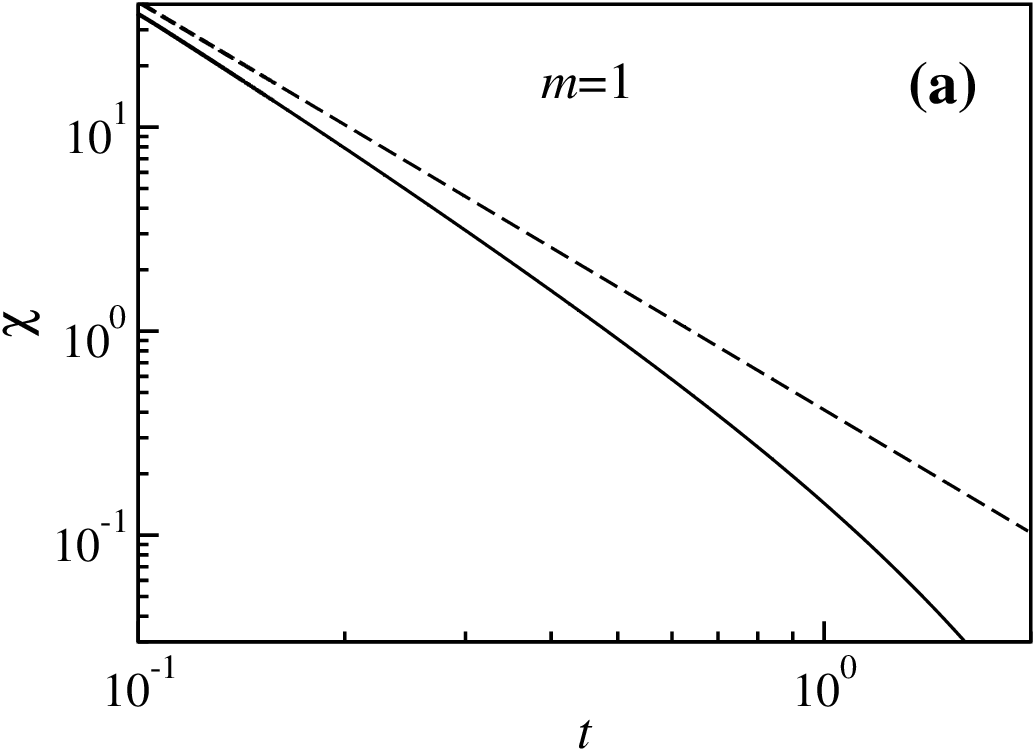}
\\[5pt]
\includegraphics[scale=0.4]{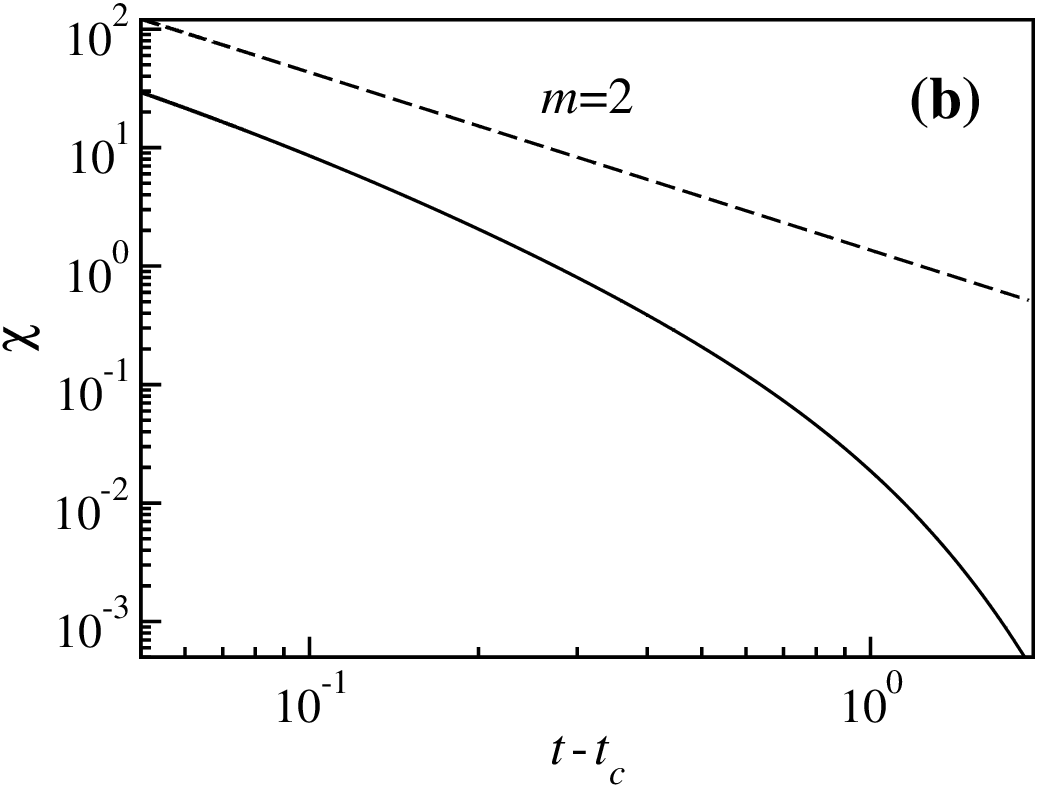}
\\[5pt]
\includegraphics[scale=0.4]{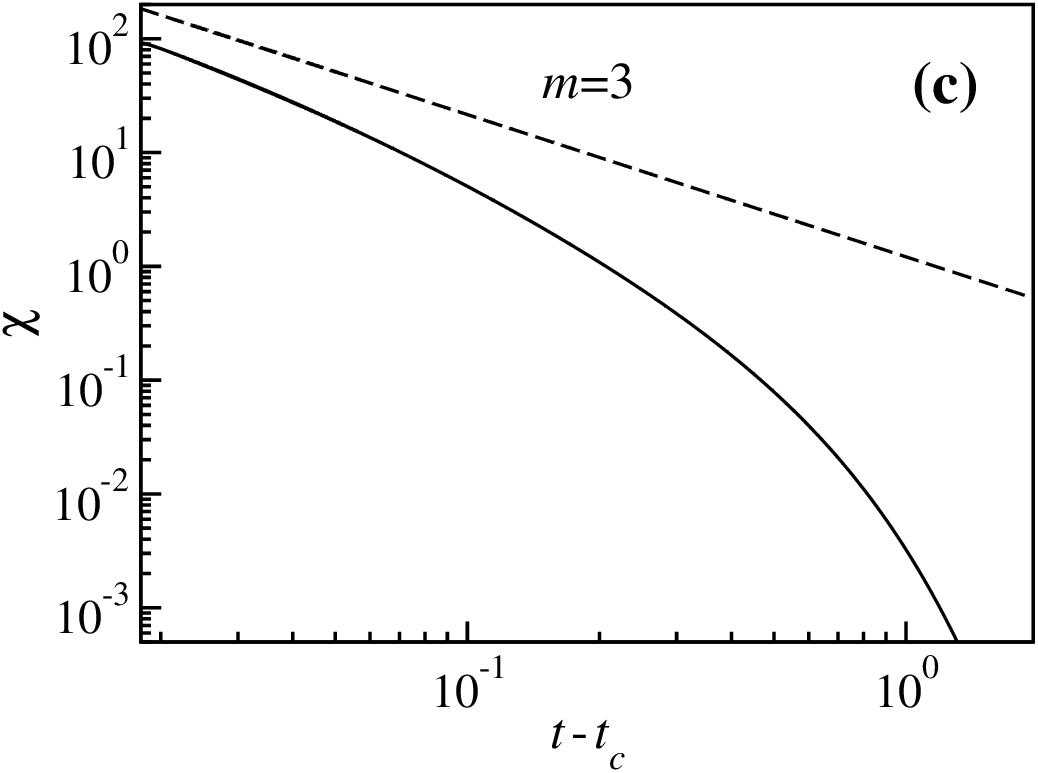}
\end{center}
\caption{
Critical behavior of the susceptibility $\chi$ for (a) $m=1$, (b) $m=2$, and (c) $m=3$.
Solid lines, numerical solution of $10^5$ evolution equations with initial condition $P(s,0)=\zeta[2m/(2m{-}1)]^{-1} s^{-2m/(2m{-}1)}$.
Dashed lines, (a) power-law in Eq.~(\ref{e4800}); (b) and (c) power-law in Eq.~(\ref{e9300}) assuming $\mu=1/(2m-2)$, and $C=0.024$ and $0.0088$ for $m=2$ and $3$, respectively.
}
\label{f5}       
\end{figure}

Figure~\ref{f5} presents the susceptibility of the infinite-order percolation transitions for $\tilde{\tau}=2+1/(2m{-}1)$. 
We obtain the solid curves in this figure inserting the numerical solution of $10^5$ evolution equations into the expression $\chi=\sum_{s} s Q(s,t)^2/P(s,t)$. The dashed lines are the power laws in Eq.~(\ref{e4800}) and Eq.~(\ref{e9300}) for $m=1$ and $m>1$, respectively. For $m=2$ and $3$ we used the value of $\mu=1/(2m-2)$ corresponding to the asymptotic distribution $P(s,t_c) \sim s^{-2m/(2m-1)} \ln s$, which is observed in Fig.~\ref{f4}.


\section{Conclusions}
\label{conclusions}

In this article we have explored the impact of 
the initial cluster size distribution in a set of models generalizing ordinary percolation. Specifically, we focused on explosive percolation models, but our approach could be applied to a much wider range of generalized percolation models that can be reduced to various aggregation processes. In particular, we considered initial cluster size distributions, for which
 the percolation phase transition
 turned out to be remarkably different from that for the evolution started from isolated nodes \cite{da2010explosive,da2014solution,da2014critical}. Our results are 
summarized 
 in Table~\ref{t1}. We have found the special values of the exponent $\tilde{\tau}$ of the initial cluster size distribution $s^{-\tilde{\tau}}$, namely $\tilde{\tau}=2+1/(2m-1)$, for which the percolation cluster emerges continuously, with all derivatives zero. For ordinary percolation, $m=1$, this singularity is at $t=0$, and the susceptibility diverges as $t^{-2}$ in contrast to the Curie--Weiss law typical for mean-field theories including various percolation problems, in which the evolution starts from isolated nodes. We have found that for explosive percolation, $m>1$, the situation is even more interesting. When $\tilde{\tau}=2+1/(2m-1)$, 
 (i) the phase transition occurs at $t_c>0$;  
 (ii) the transition is continuous, of infinite order; 
 (iii) the phase $0\leq t\leq t_c$ is critical in the sense that the generalized susceptibility diverges at any $t\leq t_c$, and the size distribution of clusters has the same asymptotics $s^{-\tilde{\tau}}$ in the entire critical phase; 
 (iv) the susceptibility diverges above the transition with a critical exponent different from $1$; 
 (v) the size distribution of clusters at $t_c$ is the same power law but with additional logarithmic factor, see Table~\ref{t1}. Finally, in the special case of $\tilde{\tau}=2+1/(2m-1)$, we have obtained unusual scaling both for ordinary and explosive percolation. 
Note that, counterintuitively, in the case of explosive percolation, for this special value of $\tilde{\tau}$ the power-law critical singularity of the generalized susceptibility is accompanied by a strong divergence of the moments $\langle s \rangle_P$ and $\langle s \rangle_Q$ at the critical point, $\langle s \rangle_P \sim \langle s \rangle_Q^2 \sim \exp [\text{const}(t-t_c)^{-\mu}]$, see Appendix~\ref{a600}. We studied the range $\tilde{\tau}>2$, in which the initial distribution $P(s,0) \sim s^{1-\tilde{\tau}}$ is normalizable and the average cluster size is finite. In the case of $\tilde{\tau} \leq 2$ and $m=1$ it was found that $S(t>0)=1$ \cite{cho2010cluster}.  

The infinite order singularity for the percolation cluster at $t=0$ was found in Ref.~\cite{leyvraz2012scaling,cho2010cluster} in aggregation processes with power-law kernels \cite{krapivsky2010kinetic,manna2011a}  instead of our power-law initial distribution of clusters for ordinary percolation. This singularity at zero was also observed in epidemic models and percolation problems on equilibrium scale-free networks with the degree distribution $P(q) \sim q^{-3}$ \cite{dorogovtsev2008critical,pastor2001epidemic,cohen2003structural}. 

The critical phase and the infinite-order phase transition resemble the Berezinskii-Kosterlitz-Thouless transition \cite{berezinskii1971destruction,kosterlitz1973ordering}, which was observed in numerous systems at a lower critical dimension. In addition, these singularities were observed in heterogeneous one-dimensional systems with long-range interactions \cite{costin1990infinite,bundaru1999phase}, in various growing networks \cite{callaway2001randomly,dorogovtsev2001anomalous,kim2002infinite,dorogovtsev2003evolution,bauer2005phase,khajeh2007berezinskii}, and in percolation on hierarchical and nonamenable graphs \cite{singh2014explosive,nogawa2014transition,hasegawa2014critical}. Interestingly, in these growing networks, the critical distributions of finite cluster sizes also had factors with powers of logarithms \cite{dorogovtsev2001anomalous,kim2002infinite}. 

In conclusion, we have found that initial conditions can have dramatic effect on the scenario of percolation transition and its various generalizations. For particular initial distributions of clusters in the case of explosive percolation, we revealed a continuous phase transition of infinite-order singularity, which resembles the Berezinskii-Kosterlitz-Thouless transition. We suggest that our findings are valid for a wide range of generalizations of percolation, in particular, for explosive percolation models with power-law kernels.




\acknowledgments


This work was partially supported by the FET proactive IP project MULTIPLEX 317532, the FCT project EXPL/FIS-NAN/1275/2013, and the project ``New Strategies Applied to Neuropathological Disorders'' (CENTRO-07-ST24-FEDER-002034) cofunded by QREN and EU.


\appendix


\section{Calculation of the coefficient $A_2(s)$ for $m=1$ and $\tilde{\tau}<3$}
\label{a400}

Let us derive expression (\ref{e2100}) for the coefficient $A_2(s)$ of the expansion (\ref{e900}) of $P(s,t)$ in the case of ordinary percolation.  
We differentiate both sides of Eq.~(\ref{e100}) with respect to $t$  
and replace $\partial_t^{(i)} P(s,t)|_{t=0}$ with $A_i(s)\, i!$. In this way we get
\begin{equation}
 A_2(s)
=  s \sum_{u=1}^{s-1} A_1(u) A_0(s{-}u) -  s A_1(s)
.
\label{e1700}
\end{equation}
Rearranging this equation 
in order to cancel divergencies, we write
\begin{align}
 A_2(s)
&=  s  \sum_{u=1}^{s-1} \left[A_1(u) -  A_1(s)\right] \left[A_0(s{-}u) - A_0(s) \right] 
\nonumber
\\[5pt]
&-s(s-1)A_0(s)A_1(s)-sA_1(s)\sum_{u=s}^{\infty} A_0(u)
\nonumber
\\[5pt]
& +sA_0(s)\sum_{u=1}^{s-1} A_1(u)
.
\label{e1800}
\end{align}
Similarly to $A_1$, 
let us  
replace the sums over $u$ in Eq.~(\ref{e1800}) by convergent integrals. 
The first and second sums on the right-hand site of this relation can be directly replaced by the respective integrals. 
So we have to analyze only the third sum.  
For this sum, there are three possibilities:  
(i) if $\tilde{\tau}<5/2$, then the asymptotics of $A_1(u)$ have the exponent $4-2\tilde{\tau}>-1$, (ii) if  $\tilde{\tau}>5/2$, then the exponent $4-2\tilde{\tau}<-1$, and (iii) if $\tilde{\tau}=5/2$, then the coefficient $a_1$ from Eq.~(\ref{e1300}) is zero, and 
$A_1(s)$ decays as $s^{-3/2}$, see Eq.~(\ref{e1500}).
In the 
the first case the sum $\sum_{u{<}s} A_1(u)$ can be directly replaced by an integral convergent at the lower limit. In the 
second and third cases this sum must be first replaced with $\sum_{u=1}^\infty A_1(u) - \sum_{u\geq s} A_1(u)$. The sum $\sum_{u=1}^\infty A_1(u)$ is a finite constant and $\sum_{u\geq s} A_1(u)$ can be replaced by a convergent integral.  
Substituting the asymptotics of $A_0$ and $A_1$, we obtain 
\begin{align}
& A_2(s)
\cong a_0 a_1  s^{7-3\tilde{\tau}}  \int_{0}^{1}dx\, \left[x^{4-2\tilde{\tau}} -  1\right] \left[(1{-}x)^{1-\tilde{\tau}} - 1\right] 
\nonumber
\\[5pt]
&-a_0 a_1s^{7-3\tilde{\tau}}-a_0 a_1s^{7-3\tilde{\tau}}\int_{1}^{\infty}dx\, x^{1-\tilde{\tau}}
\nonumber
\\[5pt]
& +\theta(5/2{-}\tilde{\tau}) a_0 a_1 s^{7-3\tilde{\tau}}\int_{0}^{1} dx\,x^{4-2\tilde{\tau}} 
\nonumber
\\[5pt]
&+\theta(\tilde{\tau}{-}5/2) a_0s^{2-\tilde{\tau}}\Bigg[ \sum_{u=1}^{\infty} A_1(u) 
-  a_1 s^{5{-}2\tilde{\tau}}\!\!\int_{1}^{\infty}\!\!\!\!dx\,  x^{4-2\tilde{\tau}} \Bigg]
,
\label{e1900}
\end{align}
where the step function $\theta(x)$ is defined here as  $\theta(x{<}0){=}0$ and $\theta(x{\geq}0){=}1$.  
When $\tilde{\tau}{>}5/2$, the fourth term on the right-hand side of this equation 
is zero, and in the last term, the sum $\sum_{u} A_1(u)=0$, see Appendix~\ref{a300}.   
When $\tilde{\tau}{=}5/2$, the coefficient $a_1{=}0$, and so the asymptotics of $A_2(s)$ is given 
$a_0s^{2-\tilde{\tau}} \sum_{u} A_1(u)$. As we show in Appendix~\ref{a300}, in this special case the sum $\sum_{u} A_1(u)$ is finite, see Eq.~(\ref{e3701}).  
As a result, for any $\tilde{\tau}<3$, we have 
\begin{equation}
A_2(s)
\cong a_0^3 \frac{2 \Gamma(2{-}\tilde{\tau})^3}{3 \Gamma(6{-}3\tilde{\tau})}
s^{7-3\tilde{\tau}} 
. 
\label{e2100a}
\end{equation}


\section{Generating functions 
approach for ordinary percolation ($m=1$)}
\label{a300}

Using generating functions, we will obtain $\sum_s A_n(s)$ for different exponents $\tilde{\tau}$ and the general expression for the coefficients $a_n$.
Let us consider the generating functions 
of the coefficients $A_n(s)$ in the 
series~(\ref{e900}), 
that is 
\begin{equation}
\rho_n(z)=\sum_{s} z^s A_n(s)
.
\label{e3300}
\end{equation}

We obtain the generating function $\rho_1(z)$ 
in terms of $\rho_0(z)$ multiplying both sides of the evolution equation~(\ref{e1100}) by $z^s$ and summing over $s$, 
\begin{eqnarray}
\rho_1(z)
&&
=2  \sum_{s}  \left[ \sum_{u=1}^{s-1} z^u u A_0(u) z^{s{-}u} A_0(s{-}u) - z^{s} s A_0(s) 
\right]
\nonumber
\\[5pt]
&&
=2 \sum_{s}  z^s s A_0(s) \left[  \sum_{u} z^u A_0(u)  - 1 \right]
\nonumber
\\[5pt]
&&
= \partial_{\ln z} (\rho_0(z) -1)^2
.
\label{e3400}
\end{eqnarray}
Here we used  
$\sum_s  \sum_{u<s} f(u) f(s{-}u) = \sum_s f(s) \sum_{u} f(u)$.

The function $\rho_0(z)$ is the generating function of the initial distribution $P(s,0)\equiv A_0(s)$. For a power-law $P(s,0)\cong a_0 s^{1-\tilde{\tau}}$ the singular behavior  of $\rho_0(z)$ at $z=1$ 
 is given by Eq.~(\ref{e400}), which we reproduce here for the sake of clarity, 
\begin{equation}
1-\rho_0(z) \cong -a_0 \Gamma(2-\tilde{\tau})(-\ln z)^{\tilde{\tau}-2}
.
\label{e3410}
\end{equation}
Inserting this result into Eq.~(\ref{e3400}) we find the singularity of $\rho_1(z)$ at $z=1$: 
\begin{equation}
\rho_1(z)
\cong - 2 a_0^2 \Gamma(2-\tilde{\tau})^2 (\tilde{\tau}-2) (1-z)^{2\tilde{\tau}-5}
.
\label{e3500}
\end{equation}
Differentiating both sides of the evolution equation with respect to $t$, and combining with Eq.~(\ref{e3400}) and~(\ref{e3410}),
we express the function $\rho_2(z)$ as
\begin{eqnarray}
\rho_2(z)
=&& \sum_s z^{s}A_0(s) \sum_{u} z^u u A_1(u)
\nonumber
\\[5pt]
&& + \sum_s z^{s} A_1(s) \sum_{u} z^u u A_0(u) - \sum_s z^s s A_1(s)
\nonumber
\\[5pt]
=&&\frac{2}{3} \partial_{\ln z} \left[ \partial_{\ln z} (\rho_0(z) -1)^3  \right]
\nonumber
\\[8pt]
\cong&& 2 a_0^3 \Gamma(2-\tilde{\tau})^3 (\tilde{\tau}-2)(3\tilde{\tau}-7) (1-z)^{3\tilde{\tau}-8}
.
\label{e3600}
\end{eqnarray}
In a similar way, for a general $n$, we obtain
\begin{eqnarray}
\rho_n(z) \cong&& \frac{2^n a_0^{n+1} \Gamma(2-\tilde{\tau})^{n+1} \Gamma[(n+1)(3-\tilde{\tau})-1] }{(n+1)! \,\Gamma[(n+1)(2-\tilde{\tau})]} 
\nonumber
\\[5pt]
&&\times (1-z)^{(n+1)(\tilde{\tau}-3)+1}
.
\label{e3700}
\end{eqnarray}
The sums $B_n \equiv \sum_s A_n(s)$ are equal to the value of the generating function $\rho_n(z)$ at $z=1$. Then according to the last equation 
\begin{equation}
B_n =
 \begin{cases}
 0 &\mbox{if } n<\beta
 ,
 \\[3pt]
 \displaystyle{-\frac{2^{n} [-a_0 \Gamma(2-\tilde{\tau})]^{n+1}   }{n+1} }
  &\mbox{if } n=\beta
  ,
\\[3pt]
\infty &\mbox{if } n>\beta
, 
\end{cases}
\label{e3701}
\end{equation}
where 
$\beta=(\tilde{\tau}-2)/(3-\tilde{\tau})$ 
is defined by the condition ${1-\tilde{\tau} + \beta(3-\tilde{\tau})=-1}$.

Equation~(\ref{e3700}) can be used to find the general form of the prefactors $a_n$ in the asymptotics of $A_n(s)$.
 We obtain the singular contribution to $\rho_n(z)$ near $z=1$ by inserting the respective power-law $A_n(s) \cong a_n s^{1-\tilde{\tau}+n\sigma}$ into Eq.~(\ref{e3300}):
\begin{equation}
\rho_n(z) \cong a_n \Gamma[(n+1)(3-\tilde{\tau})-1]  (1-z)^{(n+1)(\tilde{\tau}-3)+1}
.
\label{e3710}
\end{equation}
Comparing Eqs.~(\ref{e3710}) and~(\ref{e3700}) we readily get:
\begin{equation}
a_n= \frac{\left[2 a_0 \Gamma(2-\tilde{\tau})\right]^{n+1}}{2 \Gamma[(n+1)(2-\tilde{\tau})](n+1)!} 
.
\label{e3730}
\end{equation}


\section{Calculation of 
coefficients $A_n(s)$ for $m=1$ and $\tilde{\tau}=3$}
\label{a500}

For the particular case of $m=1$ and $\tilde{\tau}=3$ we will calculate the asymptotics of the coefficients $A_n(s)$, which differ from those 
in Sec.~\ref{2a} for $\tilde{\tau}<3$.
The difference follows from the divergence of the convolution integrals of Eqs.~(\ref{e1300}),~(\ref{e1900}), etc, for the coefficients $A_1(s)$, $A_2(s)$, etc., when $\tilde{\tau}=3$. The arguments of these integrals diverge at the upper and lower limits as $(1{-}x)^{-1}$ and  $x^{-1}$, respectively. To remove these divergences we must subtract one extra term to each factor of the convolution argument.
For $A_1$, instead of Eq.~(\ref{e1200}) and~(\ref{e1300}), we write
\begin{eqnarray}
A_1(s)& = & s \sum_{u=1}^{s-1}\left[A_0(u)-a_0 s^{-2}-2 a_0 s^{-3} (s-u)\right]
\nonumber
\\[5pt]
&\times& \left[A_0(s-u)-a_0 s^{-2} -2 a_0 s^{-3} u \right] 
\nonumber
\\[5pt]
&-& 4 a_0^2 s^{-5}\sum_{u=1}^{s-1} u(s-u) - a_0^2 (s-1)  s^{-3}
\nonumber
\\[5pt]
&-&4 a_0^2 s^{-4} \sum_{u=1}^{s-1} u+4 a_0 s^{-2} \sum_{u=1}^{s-1} u A_0(u)
\nonumber
\\[5pt]
&+& 2 a_0 s^{-1}\sum_{u=1}^{s-1} A_0(u) -2  sA_0(s)
\nonumber
\\[5pt]
&\cong&   a_0^2 s^{-2} \int_0^{1} dx \big\{  \left[x^{-2}-1 - 2 (1-x)\right]  
\nonumber
\\[5pt]
&\times&
  \left[(1-x)^{-2}-1 -2 x \right]  -4x(1-x)   -4x \big\}  - a_0^2  s^{-2} 
\nonumber
\\[5pt]
&-& 2 a_0^2 s^{-2}\int_{1}^{\infty}dx\, x^{-2} +4 a_0^2 s^{-2} \sum_{u=1}^{s-1} u^{-1}
.
\label{e4900}
\end{eqnarray}
Since  $\sum_{u<s} u^{-1}\cong \ln s$ the last term decays as $s^{-2}\ln s$, while the others decay as $s^{-2}$. Then, for large enough $s$, we obtain 
 \begin{equation}
A_1(s) \cong 4 a_0^2 s^{-2} \ln s
.
\label{e5000}
\end{equation} 
We obtain the asymptotics of the other coefficients $A_n(s)$ in a similar way. The $A_n$ found by this procedure have the general expression 
  \begin{equation}
A_n(s) \cong 2^n (n{+}1) a_0^{n+1} s^{-2} \left(\ln s\right)^n
.
\label{e5101}
\end{equation}


\section{Expansion of $P(s,t)$ at $t_c>0$ for $m>1$ and $\tilde{\tau}=2+1/(2m{-}1)$}
\label{a100}

Let us obtain the scaling form of $P(s,t)$ near the critical point $t_c>0$ 
in the case of $m>1$ and $\tilde{\tau}=2+1/(2m{-}1)$. 
Our numerical solution of exact evolution equations 
shows that 
that asymptotics of the distribution $P(s,t_c)$ differs from a pure power-law by a factor $\ln^\lambda s$ with $\lambda$ close to $1$, see Fig.~\ref{f4} and Appendix~\ref{a200}. 
Then, at $t_c$ we have
\begin{equation}
P(s,t_c)\cong a_0 C (\ln s)^\lambda s^{-2m/(2m-1)} 
,
\label{se8810}
\end{equation}
where $C$ is a constant.

By the same method as in the Secs.~\ref{2a}, \ref{2b}, \ref{3a}, and \ref{3b}, 
we obtain the expansion of $P(s,t)$ around $t_c$ 
starting from the distribution $P(s,t_c)$ in Eq.~(\ref{se8810}). 
The resulting asymptotics of the coefficient $A_n$ is
 \begin{equation}
 A_n(s)\cong a_n [C (\ln s)^{\lambda}]^{1+n(2m-1)}  s^{-2m/(2m-1)}
\label{e8820}
\end{equation}
with prefactors $a_n$ given by Eq.~(\ref{e7710}).
The series with these coefficients gives
\begin{equation}
P(s,t)\cong C (\ln s)^\lambda s^{-2m/(2m-1)} f\left[(t-t_c)\{C (\ln s)^\lambda\}^{2m-1}\right]
,
\label{e8840e}
\end{equation}
where $f(x)$ is the function from Eq.~(\ref{e7800}). 
This scaling form is 
similar to Eq.~(\ref{e5200}) for $m=1$, $\tilde{\tau}=3$, that is, a logarithmic factor of $s$ appears in the argument of the scaling function. The difference, however, is that for $m=1$ we have $t_c=0$, while for $m>1$ we have $t_c>0$.


\section{Percolation cluster size singularity at $t_c>0$ for $m>1$ and $\tilde{\tau}=2+1/(2m{-}1)$}
\label{a200}

In this appendix, we show for $m>1$ and $\tilde{\tau}=2+1/(2m{-}1)$ that an infinite-order percolation transition takes place at the critical point $t_c=r$ given by Eq.~(\ref{e7810}). When $t<t_c$, 
the percolation cluster is absent, $S=0$. 
Let us obtain the critical singularity of $S(t>t_c)$  and relate it 
with the asymptotics of $P(s,t_c)$.

Figure~\ref{f4} shows that at $t_c$ the asymptotics of $P(s,t_c)$ differs from the initial power-law one by a factor
\begin{equation}
H(s) \equiv \frac{P(s,t_c)}{a_0s^{-2m/(2m-1)} }
.
\label{se8210}
\end{equation}
For the sake of brevity we do not indicate that the function $H(s)$ depends on $m$. 
Substituting $P(s,t_c)=H(s) a_0 s^{-2m/(2m-1)}$ into Eq.~(\ref{e400}) (this equation is valid for any $m$) we find the asymptotic function near $z=1$ at the critical point,  i.e., $\rho(z,t_c)=\sum_s  P(s,t_c) z^s$ 
\begin{eqnarray}
&&1-\rho(z,t_c)
\cong a_0 \sum_s (1-z^s) H(s) s^{-2m/(2m-1)}
\nonumber
\\[5pt]
&&\cong a_0 (- \ln z)^{1/(2m-1)}\!\! \int_0^\infty\!\!\! dx\, (1{-}e^{-x}) H\!\! \left(\! \frac{-x}{\ln z} \! \right)\! x^{-2m/(2m{-}1)}
\nonumber
\\&&\cong - a_0 \Gamma[-1/(2m-1)] H\left( \frac{-1}{\ln z} \right) \left(- \ln z \right)^{1/(2m-1)}
.
\label{e8400}
\end{eqnarray}
Here we assumed that $H(s)$ grows with $s$ slower than any power law, which means that $H(-x / \ln z) / H(-1/\ln z) \to 1$ when $z \to 1$.

The inverse function of the singularity of $\rho(z,t_c)$ is the
the initial condition for solving the differential equation~(\ref{e5800}) in the phase with $S>0$. 
Inverting Eq.~(\ref{e8400}) we get
\begin{equation}
\ln z \cong - \left(\frac{\rho-1}{a_0\Gamma[-1/(2m-1)] H\left[(1-\rho)^{1-2m}\right]}\right)^{2m-1}
\label{e8500}
\end{equation}
at $t=t_c$. 
With this initial condition, the solution of Eq.~(\ref{e5800}) at $z=1$ is
 \begin{eqnarray}
\!\!\!\!\!\!\!\!\!\!\!\! && \left(\frac{-S(t)}{a_0 \Gamma[-1/(2m{-}1)] H\left[S(t)^{1-2m}\right]}\right)^{2m-1}=
\nonumber
\\[5pt]
\!\!\!\!\!\!\!\!\!\!\!\!
&&2  m^2S(t)\!\!\int_{t_c}^{t}\!\! dt' S(t')^{2m-2} \!\!- 2 (m^2{-}m)\!\! \int_{t_c}^{t}\!\! dt' S(t')^{2m-1}
\label{e8600}
\end{eqnarray}
for $t$ approaching $t_c$ from above.

Let us introduce the function
\begin{equation}
D(t)\equiv - \ln S(t)
.
\label{e8510}
\end{equation}
For small $t-t_c$ the integrals in Eq.~(\ref{e8600}) can be calculated by changing the integration variable to $nD(t')\equiv x$, where $n=2m-2$ and $2m-1$ for the first and the second integral on the right-hand side of Eq.~(\ref{e8600}), respectively, 
 \begin{eqnarray}
\int_{t_c}^{t}\ dt'\, S(t')^n &&= \int_{t_c}^{t} dt'\, \exp[ -nD(t') ] 
\nonumber
\\[5pt]
&&= -\int_{nD(t)}^\infty  dx\, \left( \frac{\partial x}{\partial t'}\right)^{-1} \exp( -x ) 
\nonumber
\\[5pt]
&&\cong - \left( n \frac{\partial D(t)}{\partial t}\right)^{-1} \int_{nD(t)}^\infty  dx\, \exp( -x ) 
\nonumber
\\[5pt]
&&= - \left( n \frac{\partial D(t)}{\partial t}\right)^{-1} S(t)^n
. 
\label{e8610}
\end{eqnarray}
Notice that the lower limit of the integral in this equation $nD(t) \to \infty$ when $t$ approaches $t_c$.
Substituting Eq.~(\ref{e8610}) into Eq.~(\ref{e8600}) we obtain 
 \begin{equation}
 \frac{H\left[S(t)^{1-2m}\right]^{2m-1}}{\{-a_0 \Gamma[-1/(2m{-}1)] \}^{1-2m}}
 = -\frac{(2m{-}1)(m{-}1)}{m(3m-2)} \frac{\partial D(t)}{\partial t}
.
\label{e8630}
\end{equation}

Our numerical solution of exact evolution equations, Fig.~\ref{f4}, shows that at the critical point, $P(s,t_c) \sim s^{1-\tilde{\tau}}\ln^\lambda s$, where the exponent $\lambda$ was found to be close to $1$. This asymptotics implies that 
\begin{equation}
H(x) \cong C \left( \ln x \right)^{\lambda}
,
\label{e8660}
\end{equation}
for large $x$, where $C$ is a constant. 
Inserting this expression into Eq.~(\ref{e8630}) we obtain the differential equation
 \begin{equation}
\frac{\partial D}{\partial t}{=}{-}\frac{m(3m{-}2)\{-a_0 C \Gamma[-1/(2m{-}1)]\}^{2m-1}}{(m{-}1)(2m{-}1)^{1-\lambda(2m-1)}}D^{\lambda(2m-1)}
,
\label{e8670}
\end{equation}
for ${-}\ln S\equiv D$. With the initial condition $S(t_c)=0$ the solution of this equation is 
 \begin{equation}
S \sim \exp(-d \delta^{-\mu})
,
\label{e8700}
\end{equation}
where 
\begin{equation}
\mu=\frac{1}{\lambda(2m-1)-1}
,
\label{e8710}
\end{equation}
 and $d$ is a constant
 \begin{equation}
 {}\!\!\!
d = \frac{1}{2m-1}  \left(  \frac{m(3m-2) (m-1)^{-1}}{
\mu  \{-a_0 C \Gamma[-1/(2m-1)]\}^{1-2m}}\right)^{-\mu}
\!\!.
\label{e8800}
\end{equation}
The values of $\mu$ from $0$ to $\infty$ corresponds to exponent $\lambda$, respectively, from $\infty$ to $1/(2m{-}1)$. The latter is the lower bound for $\lambda$.


\section{First moments of distributions $P(s,t)$ and $Q(s,t)$}
\label{a600}

In this section we find the singularities of the first moments of the distributions $P(s,t)$ and $Q(s,t)$, $\langle s \rangle_P$ and $\langle s \rangle_Q$, respectively, using 
the results of the previous sections. 
For $m=1$ the distribution $Q(s,t)=P(s,t)$, and so $\langle s \rangle_P=\langle s \rangle_Q=\chi$. In the region $\tilde{\tau}<3$, according to Eq.~({\ref{e4700}), we have
\begin{equation}
\langle s \rangle_P=\langle s \rangle_Q \cong \frac{\tilde{\tau}-2}{2(3-\tilde{\tau})} t^{-1}
,
\label{e9410}
\end{equation}
while for $\tilde{\tau}=3$ Eq.~({\ref{e4800}) gives
\begin{equation}
\langle s \rangle_P=\langle s \rangle_Q \cong \frac{1}{4 a_0} t^{-2}
.
\label{e9420}
\end{equation}

For $m>1$ we find the singularities of $\langle s \rangle_P$ and $\langle s \rangle_Q$ inserting the previously obtained singularity of the relative size of the percolation cluster $S$ into Eqs.~(\ref{e9000}) and~(\ref{e8200}), respectively.
 For $\tilde{\tau}<2{+}1/(2m{-}1)$ the power law $S\cong B t^\beta$, Eq.~(\ref{e5901}),
 results in the first moments
\begin{equation}
\langle s \rangle_P \cong  \frac{B^{2(1-m)}}{2m} t^{2(1-m)\beta-1}
\label{e9500}
\end{equation} 
and
\begin{equation}
\langle s \rangle_Q \cong \frac{B^{1-m}}{2} t^{(1-m)\beta-1}
,
\label{e9400}
\end{equation} 
where $\beta$ and $B$ are given by Eqs.~(\ref{e6000}) and~(\ref{e6100}), respectively. 
For $m>1$ and $\tilde{\tau}=2{+}1/(2m{-}1)$, the infinite-order transition occurs at $t_c>0$ with the singularity $S\sim \exp(-d \delta^{-\mu})$. This critical behavior leads to the following singularities of the first moments:  
\begin{equation}
\langle s \rangle_P \sim \exp[2d(m-1) \delta^{-\mu}]
\label{e9600}
\end{equation} 
and
\begin{equation}
\langle s \rangle_Q \sim \exp[d(m-1) \delta^{-\mu}]
\label{e9700}
\end{equation} 
for $t>t_c$,
where the exponent $\mu$ and the constant $d$ are given by Eqs.~(\ref{e8835}) and~(\ref{e8837}), respectively. 
In this situation the first moments $\langle s \rangle_P$ and $\langle s \rangle_Q$ exhibit exponential divergences approaching $t_c$ from above.
In the ``critical phase'', $t<t_c$, the moments  $\langle s \rangle_P$ and $\langle s \rangle_Q$ and the susceptibility $\chi$ are divergent, see Sec.~\ref{3b}.


\bibliographystyle{apsrev4-1}
\bibliography{refs}

\begin{thebibliography}{42}%
\makeatletter
\providecommand \@ifxundefined [1]{%
 \@ifx{#1\undefined}
}%
\providecommand \@ifnum [1]{%
 \ifnum #1\expandafter \@firstoftwo
 \else \expandafter \@secondoftwo
 \fi
}%
\providecommand \@ifx [1]{%
 \ifx #1\expandafter \@firstoftwo
 \else \expandafter \@secondoftwo
 \fi
}%
\providecommand \natexlab [1]{#1}%
\providecommand \enquote  [1]{``#1''}%
\providecommand \bibnamefont  [1]{#1}%
\providecommand \bibfnamefont [1]{#1}%
\providecommand \citenamefont [1]{#1}%
\providecommand \href@noop [0]{\@secondoftwo}%
\providecommand \href [0]{\begingroup \@sanitize@url \@href}%
\providecommand \@href[1]{\@@startlink{#1}\@@href}%
\providecommand \@@href[1]{\endgroup#1\@@endlink}%
\providecommand \@sanitize@url [0]{\catcode `\\12\catcode `\$12\catcode
  `\&12\catcode `\#12\catcode `\^12\catcode `\_12\catcode `\%12\relax}%
\providecommand \@@startlink[1]{}%
\providecommand \@@endlink[0]{}%
\providecommand \url  [0]{\begingroup\@sanitize@url \@url }%
\providecommand \@url [1]{\endgroup\@href {#1}{\urlprefix }}%
\providecommand \urlprefix  [0]{URL }%
\providecommand \Eprint [0]{\href }%
\providecommand \doibase [0]{http://dx.doi.org/}%
\providecommand \selectlanguage [0]{\@gobble}%
\providecommand \bibinfo  [0]{\@secondoftwo}%
\providecommand \bibfield  [0]{\@secondoftwo}%
\providecommand \translation [1]{[#1]}%
\providecommand \BibitemOpen [0]{}%
\providecommand \bibitemStop [0]{}%
\providecommand \bibitemNoStop [0]{.\EOS\space}%
\providecommand \EOS [0]{\spacefactor3000\relax}%
\providecommand \BibitemShut  [1]{\csname bibitem#1\endcsname}%
\let\auto@bib@innerbib\@empty
\bibitem [{\citenamefont {Stauffer}(1979)}]{stauffer1979scaling}%
  \BibitemOpen
  \bibfield  {author} {\bibinfo {author} {\bibfnamefont {D.}~\bibnamefont
  {Stauffer}},\ }\href@noop {} {\bibfield  {journal} {\bibinfo  {journal}
  {Phys. Rep.}\ }\textbf {\bibinfo {volume} {54}},\ \bibinfo {pages} {1}
  (\bibinfo {year} {1979})}\BibitemShut {NoStop}%
\bibitem [{\citenamefont {Stauffer}\ and\ \citenamefont
  {Aharony}(1991)}]{stauffer1991introduction}%
  \BibitemOpen
  \bibfield  {author} {\bibinfo {author} {\bibfnamefont {D.}~\bibnamefont
  {Stauffer}}\ and\ \bibinfo {author} {\bibfnamefont {A.}~\bibnamefont
  {Aharony}},\ }\href@noop {} {\emph {\bibinfo {title} {{Introduction to
  Percolation Theory}}}}\ (\bibinfo  {publisher} {Taylor and Francis},\
  \bibinfo {address} {London},\ \bibinfo {year} {1991})\BibitemShut {NoStop}%
\bibitem [{\citenamefont {Dorogovtsev}\ \emph {et~al.}(2008)\citenamefont
  {Dorogovtsev}, \citenamefont {Goltsev},\ and\ \citenamefont
  {Mendes}}]{dorogovtsev2008critical}%
  \BibitemOpen
  \bibfield  {author} {\bibinfo {author} {\bibfnamefont {S.~N.}\ \bibnamefont
  {Dorogovtsev}}, \bibinfo {author} {\bibfnamefont {A.~V.}\ \bibnamefont
  {Goltsev}}, \ and\ \bibinfo {author} {\bibfnamefont {J.~F.~F.}\ \bibnamefont
  {Mendes}},\ }\href@noop {} {\bibfield  {journal} {\bibinfo  {journal} {Rev.
  Mod. Phys.}\ }\textbf {\bibinfo {volume} {80}},\ \bibinfo {pages} {1275}
  (\bibinfo {year} {2008})}\BibitemShut {NoStop}%
\bibitem [{\citenamefont {Dorogovtsev}(2010)}]{dorogovtsev2010lectures}%
  \BibitemOpen
  \bibfield  {author} {\bibinfo {author} {\bibfnamefont {S.~N.}\ \bibnamefont
  {Dorogovtsev}},\ }\href@noop {} {\emph {\bibinfo {title} {{Lectures on
  Complex Networks}}}}\ (\bibinfo  {publisher} {Oxford University Press},\
  \bibinfo {address} {Oxford},\ \bibinfo {year} {2010})\BibitemShut {NoStop}%
\bibitem [{\citenamefont {Achlioptas}\ \emph {et~al.}(2009)\citenamefont
  {Achlioptas}, \citenamefont {D'Souza},\ and\ \citenamefont
  {Spencer}}]{achlioptas2009explosive}%
  \BibitemOpen
  \bibfield  {author} {\bibinfo {author} {\bibfnamefont {D.}~\bibnamefont
  {Achlioptas}}, \bibinfo {author} {\bibfnamefont {R.~M.}\ \bibnamefont
  {D'Souza}}, \ and\ \bibinfo {author} {\bibfnamefont {J.}~\bibnamefont
  {Spencer}},\ }\href@noop {} {\bibfield  {journal} {\bibinfo  {journal}
  {Science}\ }\textbf {\bibinfo {volume} {323}},\ \bibinfo {pages} {1453}
  (\bibinfo {year} {2009})}\BibitemShut {NoStop}%
\bibitem [{\citenamefont {Ziff}(2010)}]{ziff2010scaling}%
  \BibitemOpen
  \bibfield  {author} {\bibinfo {author} {\bibfnamefont {R.~M.}\ \bibnamefont
  {Ziff}},\ }\href@noop {} {\bibfield  {journal} {\bibinfo  {journal} {Phys.
  Rev. E}\ }\textbf {\bibinfo {volume} {82}},\ \bibinfo {pages} {051105}
  (\bibinfo {year} {2010})}\BibitemShut {NoStop}%
\bibitem [{\citenamefont {Friedman}\ and\ \citenamefont
  {Landsberg}(2009)}]{friedman2009construction}%
  \BibitemOpen
  \bibfield  {author} {\bibinfo {author} {\bibfnamefont {E.~J.}\ \bibnamefont
  {Friedman}}\ and\ \bibinfo {author} {\bibfnamefont {A.~S.}\ \bibnamefont
  {Landsberg}},\ }\href@noop {} {\bibfield  {journal} {\bibinfo  {journal}
  {Phys. Rev. Lett.}\ }\textbf {\bibinfo {volume} {103}},\ \bibinfo {pages}
  {255701} (\bibinfo {year} {2009})}\BibitemShut {NoStop}%
\bibitem [{\citenamefont {D'Souza}\ and\ \citenamefont
  {Mitzenmacher}(2010)}]{d2010local}%
  \BibitemOpen
  \bibfield  {author} {\bibinfo {author} {\bibfnamefont {R.~M.}\ \bibnamefont
  {D'Souza}}\ and\ \bibinfo {author} {\bibfnamefont {M.}~\bibnamefont
  {Mitzenmacher}},\ }\href@noop {} {\bibfield  {journal} {\bibinfo  {journal}
  {Phys. Rev. Lett.}\ }\textbf {\bibinfo {volume} {104}},\ \bibinfo {pages}
  {195702} (\bibinfo {year} {2010})}\BibitemShut {NoStop}%
\bibitem [{\citenamefont {Ziff}(2009)}]{ziff2009explosive}%
  \BibitemOpen
  \bibfield  {author} {\bibinfo {author} {\bibfnamefont {R.~M.}\ \bibnamefont
  {Ziff}},\ }\href@noop {} {\bibfield  {journal} {\bibinfo  {journal} {Phys.
  Rev. Lett.}\ }\textbf {\bibinfo {volume} {103}},\ \bibinfo {pages} {045701}
  (\bibinfo {year} {2009})}\BibitemShut {NoStop}%
\bibitem [{\citenamefont {Cho}\ \emph {et~al.}(2009)\citenamefont {Cho},
  \citenamefont {Kim}, \citenamefont {Park}, \citenamefont {Kahng},\ and\
  \citenamefont {Kim}}]{cho2009percolation}%
  \BibitemOpen
  \bibfield  {author} {\bibinfo {author} {\bibfnamefont {Y.~S.}\ \bibnamefont
  {Cho}}, \bibinfo {author} {\bibfnamefont {J.~S.}\ \bibnamefont {Kim}},
  \bibinfo {author} {\bibfnamefont {J.}~\bibnamefont {Park}}, \bibinfo {author}
  {\bibfnamefont {B.}~\bibnamefont {Kahng}}, \ and\ \bibinfo {author}
  {\bibfnamefont {D.}~\bibnamefont {Kim}},\ }\href@noop {} {\bibfield
  {journal} {\bibinfo  {journal} {Phys. Rev. Lett.}\ }\textbf {\bibinfo
  {volume} {103}},\ \bibinfo {pages} {135702} (\bibinfo {year}
  {2009})}\BibitemShut {NoStop}%
\bibitem [{\citenamefont {Radicchi}\ and\ \citenamefont
  {Fortunato}(2009)}]{radicchi2009explosive}%
  \BibitemOpen
  \bibfield  {author} {\bibinfo {author} {\bibfnamefont {F.}~\bibnamefont
  {Radicchi}}\ and\ \bibinfo {author} {\bibfnamefont {S.}~\bibnamefont
  {Fortunato}},\ }\href@noop {} {\bibfield  {journal} {\bibinfo  {journal}
  {Phys. Rev. Lett.}\ }\textbf {\bibinfo {volume} {103}},\ \bibinfo {pages}
  {168701} (\bibinfo {year} {2009})}\BibitemShut {NoStop}%
\bibitem [{\citenamefont {Radicchi}\ and\ \citenamefont
  {Fortunato}(2010)}]{radicchi2010explosive}%
  \BibitemOpen
  \bibfield  {author} {\bibinfo {author} {\bibfnamefont {F.}~\bibnamefont
  {Radicchi}}\ and\ \bibinfo {author} {\bibfnamefont {S.}~\bibnamefont
  {Fortunato}},\ }\href@noop {} {\bibfield  {journal} {\bibinfo  {journal}
  {Phys. Rev. E}\ }\textbf {\bibinfo {volume} {81}},\ \bibinfo {pages} {036110}
  (\bibinfo {year} {2010})}\BibitemShut {NoStop}%
\bibitem [{\citenamefont {Ara{\'u}jo}\ \emph {et~al.}(2011)\citenamefont
  {Ara{\'u}jo}, \citenamefont {Andrade}, \citenamefont {Ziff},\ and\
  \citenamefont {Herrmann}}]{araujo2011tricritical}%
  \BibitemOpen
  \bibfield  {author} {\bibinfo {author} {\bibfnamefont {N.~A.~M.}\
  \bibnamefont {Ara{\'u}jo}}, \bibinfo {author} {\bibfnamefont {J.~S.}\
  \bibnamefont {Andrade}}, \bibinfo {author} {\bibfnamefont {R.~M.}\
  \bibnamefont {Ziff}}, \ and\ \bibinfo {author} {\bibfnamefont {H.~J.}\
  \bibnamefont {Herrmann}},\ }\href@noop {} {\bibfield  {journal} {\bibinfo
  {journal} {Phys. Rev. Lett.}\ }\textbf {\bibinfo {volume} {106}},\ \bibinfo
  {pages} {095703} (\bibinfo {year} {2011})}\BibitemShut {NoStop}%
\bibitem [{\citenamefont {da~Costa}\ \emph {et~al.}(2010)\citenamefont
  {da~Costa}, \citenamefont {Dorogovtsev}, \citenamefont {Goltsev},\ and\
  \citenamefont {Mendes}}]{da2010explosive}%
  \BibitemOpen
  \bibfield  {author} {\bibinfo {author} {\bibfnamefont {R.~A.}\ \bibnamefont
  {da~Costa}}, \bibinfo {author} {\bibfnamefont {S.~N.}\ \bibnamefont
  {Dorogovtsev}}, \bibinfo {author} {\bibfnamefont {A.~V.}\ \bibnamefont
  {Goltsev}}, \ and\ \bibinfo {author} {\bibfnamefont {J.~F.~F.}\ \bibnamefont
  {Mendes}},\ }\href@noop {} {\bibfield  {journal} {\bibinfo  {journal} {Phys.
  Rev. Lett.}\ }\textbf {\bibinfo {volume} {105}},\ \bibinfo {pages} {255701}
  (\bibinfo {year} {2010})}\BibitemShut {NoStop}%
\bibitem [{\citenamefont {da~Costa}\ \emph
  {et~al.}(2014{\natexlab{a}})\citenamefont {da~Costa}, \citenamefont
  {Dorogovtsev}, \citenamefont {Goltsev},\ and\ \citenamefont
  {Mendes}}]{da2014critical}%
  \BibitemOpen
  \bibfield  {author} {\bibinfo {author} {\bibfnamefont {R.~A.}\ \bibnamefont
  {da~Costa}}, \bibinfo {author} {\bibfnamefont {S.~N.}\ \bibnamefont
  {Dorogovtsev}}, \bibinfo {author} {\bibfnamefont {A.~V.}\ \bibnamefont
  {Goltsev}}, \ and\ \bibinfo {author} {\bibfnamefont {J.~F.~F.}\ \bibnamefont
  {Mendes}},\ }\href@noop {} {\bibfield  {journal} {\bibinfo  {journal} {Phys.
  Rev. E}\ }\textbf {\bibinfo {volume} {89}},\ \bibinfo {pages} {042148}
  (\bibinfo {year} {2014}{\natexlab{a}})}\BibitemShut {NoStop}%
\bibitem [{\citenamefont {Nagler}\ \emph {et~al.}(2011)\citenamefont {Nagler},
  \citenamefont {Levina},\ and\ \citenamefont {Timme}}]{nagler2011impact}%
  \BibitemOpen
  \bibfield  {author} {\bibinfo {author} {\bibfnamefont {J.}~\bibnamefont
  {Nagler}}, \bibinfo {author} {\bibfnamefont {A.}~\bibnamefont {Levina}}, \
  and\ \bibinfo {author} {\bibfnamefont {M.}~\bibnamefont {Timme}},\
  }\href@noop {} {\bibfield  {journal} {\bibinfo  {journal} {Nature Phys.}\
  }\textbf {\bibinfo {volume} {7}},\ \bibinfo {pages} {265} (\bibinfo {year}
  {2011})}\BibitemShut {NoStop}%
\bibitem [{\citenamefont {Grassberger}\ \emph {et~al.}(2011)\citenamefont
  {Grassberger}, \citenamefont {Christensen}, \citenamefont {Bizhani},
  \citenamefont {Son},\ and\ \citenamefont
  {Paczuski}}]{grassberger2011explosive}%
  \BibitemOpen
  \bibfield  {author} {\bibinfo {author} {\bibfnamefont {P.}~\bibnamefont
  {Grassberger}}, \bibinfo {author} {\bibfnamefont {C.}~\bibnamefont
  {Christensen}}, \bibinfo {author} {\bibfnamefont {G.}~\bibnamefont
  {Bizhani}}, \bibinfo {author} {\bibfnamefont {S.-W.}\ \bibnamefont {Son}}, \
  and\ \bibinfo {author} {\bibfnamefont {M.}~\bibnamefont {Paczuski}},\
  }\href@noop {} {\bibfield  {journal} {\bibinfo  {journal} {Phys. Rev. Lett.}\
  }\textbf {\bibinfo {volume} {106}},\ \bibinfo {pages} {225701} (\bibinfo
  {year} {2011})}\BibitemShut {NoStop}%
\bibitem [{\citenamefont {Lee}\ \emph {et~al.}(2011)\citenamefont {Lee},
  \citenamefont {Kim},\ and\ \citenamefont {Park}}]{lee2011continuity}%
  \BibitemOpen
  \bibfield  {author} {\bibinfo {author} {\bibfnamefont {H.~K.}\ \bibnamefont
  {Lee}}, \bibinfo {author} {\bibfnamefont {B.~J.}\ \bibnamefont {Kim}}, \ and\
  \bibinfo {author} {\bibfnamefont {H.}~\bibnamefont {Park}},\ }\href@noop {}
  {\bibfield  {journal} {\bibinfo  {journal} {Phys. Rev. E}\ }\textbf {\bibinfo
  {volume} {84}},\ \bibinfo {pages} {020101(R)} (\bibinfo {year}
  {2011})}\BibitemShut {NoStop}%
\bibitem [{\citenamefont {Fortunato}\ and\ \citenamefont
  {Radicchi}(2011)}]{fortunato2011explosive}%
  \BibitemOpen
  \bibfield  {author} {\bibinfo {author} {\bibfnamefont {S.}~\bibnamefont
  {Fortunato}}\ and\ \bibinfo {author} {\bibfnamefont {F.}~\bibnamefont
  {Radicchi}},\ }\href@noop {} {\bibfield  {journal} {\bibinfo  {journal} {J.
  Phys.: Conf. Ser.}\ }\textbf {\bibinfo {volume} {297}},\ \bibinfo {pages}
  {012009} (\bibinfo {year} {2011})}\BibitemShut {NoStop}%
\bibitem [{\citenamefont {Riordan}\ and\ \citenamefont
  {Warnke}(2011)}]{riordan2011explosive}%
  \BibitemOpen
  \bibfield  {author} {\bibinfo {author} {\bibfnamefont {O.}~\bibnamefont
  {Riordan}}\ and\ \bibinfo {author} {\bibfnamefont {L.}~\bibnamefont
  {Warnke}},\ }\href@noop {} {\bibfield  {journal} {\bibinfo  {journal}
  {Science}\ }\textbf {\bibinfo {volume} {333}},\ \bibinfo {pages} {322}
  (\bibinfo {year} {2011})}\BibitemShut {NoStop}%
\bibitem [{\citenamefont {da~Costa}\ \emph
  {et~al.}(2014{\natexlab{b}})\citenamefont {da~Costa}, \citenamefont
  {Dorogovtsev}, \citenamefont {Goltsev},\ and\ \citenamefont
  {Mendes}}]{da2014solution}%
  \BibitemOpen
  \bibfield  {author} {\bibinfo {author} {\bibfnamefont {R.~A.}\ \bibnamefont
  {da~Costa}}, \bibinfo {author} {\bibfnamefont {S.~N.}\ \bibnamefont
  {Dorogovtsev}}, \bibinfo {author} {\bibfnamefont {A.~V.}\ \bibnamefont
  {Goltsev}}, \ and\ \bibinfo {author} {\bibfnamefont {J.~F.~F.}\ \bibnamefont
  {Mendes}},\ }\href@noop {} {\bibfield  {journal} {\bibinfo  {journal} {Phys.
  Rev. E}\ }\textbf {\bibinfo {volume} {90}},\ \bibinfo {pages} {022145}
  (\bibinfo {year} {2014}{\natexlab{b}})}\BibitemShut {NoStop}%
\bibitem [{\citenamefont {Cho}\ \emph {et~al.}(2010)\citenamefont {Cho},
  \citenamefont {Kahng},\ and\ \citenamefont {Kim}}]{cho2010cluster}%
  \BibitemOpen
  \bibfield  {author} {\bibinfo {author} {\bibfnamefont {Y.~S.}\ \bibnamefont
  {Cho}}, \bibinfo {author} {\bibfnamefont {B.}~\bibnamefont {Kahng}}, \ and\
  \bibinfo {author} {\bibfnamefont {D.}~\bibnamefont {Kim}},\ }\href@noop {}
  {\bibfield  {journal} {\bibinfo  {journal} {Phys. Rev. E}\ }\textbf {\bibinfo
  {volume} {81}},\ \bibinfo {pages} {030103} (\bibinfo {year}
  {2010})}\BibitemShut {NoStop}%
\bibitem [{\citenamefont {Krapivsky}\ \emph {et~al.}(2010)\citenamefont
  {Krapivsky}, \citenamefont {Redner},\ and\ \citenamefont
  {Ben-Naim}}]{krapivsky2010kinetic}%
  \BibitemOpen
  \bibfield  {author} {\bibinfo {author} {\bibfnamefont {P.~L.}\ \bibnamefont
  {Krapivsky}}, \bibinfo {author} {\bibfnamefont {S.}~\bibnamefont {Redner}}, \
  and\ \bibinfo {author} {\bibfnamefont {E.}~\bibnamefont {Ben-Naim}},\
  }\href@noop {} {\emph {\bibinfo {title} {{A Kinetic View of Statistical
  Physics}}}}\ (\bibinfo  {publisher} {Cambridge University Press},\ \bibinfo
  {address} {Cambridge},\ \bibinfo {year} {2010})\BibitemShut {NoStop}%
\bibitem [{\citenamefont {Smoluchowski}(1916)}]{smoluchowski1916brownsche}%
  \BibitemOpen
  \bibfield  {author} {\bibinfo {author} {\bibfnamefont {M.~V.}\ \bibnamefont
  {Smoluchowski}},\ }\href@noop {} {\bibfield  {journal} {\bibinfo  {journal}
  {Annalen der Physik}\ }\textbf {\bibinfo {volume} {353}},\ \bibinfo {pages}
  {1103} (\bibinfo {year} {1916})}\BibitemShut {NoStop}%
\bibitem [{\citenamefont {Nakanishi}\ and\ \citenamefont
  {Stanley}(1980)}]{nakanishi1980scaling}%
  \BibitemOpen
  \bibfield  {author} {\bibinfo {author} {\bibfnamefont {H.}~\bibnamefont
  {Nakanishi}}\ and\ \bibinfo {author} {\bibfnamefont {H.~E.}\ \bibnamefont
  {Stanley}},\ }\href@noop {} {\bibfield  {journal} {\bibinfo  {journal} {Phys.
  Rev. B}\ }\textbf {\bibinfo {volume} {22}},\ \bibinfo {pages} {2466}
  (\bibinfo {year} {1980})}\BibitemShut {NoStop}%
\bibitem [{\citenamefont {Leyvraz}(2012)}]{leyvraz2012scaling}%
  \BibitemOpen
  \bibfield  {author} {\bibinfo {author} {\bibfnamefont {F.}~\bibnamefont
  {Leyvraz}},\ }\href@noop {} {\bibfield  {journal} {\bibinfo  {journal} {J.
  Phys. A}\ }\textbf {\bibinfo {volume} {45}},\ \bibinfo {pages} {125002}
  (\bibinfo {year} {2012})}\BibitemShut {NoStop}%
\bibitem [{\citenamefont {Manna}\ and\ \citenamefont
  {Chatterjee}(2011)}]{manna2011a}%
  \BibitemOpen
  \bibfield  {author} {\bibinfo {author} {\bibfnamefont {S.~S.}\ \bibnamefont
  {Manna}}\ and\ \bibinfo {author} {\bibfnamefont {A.}~\bibnamefont
  {Chatterjee}},\ }\href@noop {} {\bibfield  {journal} {\bibinfo  {journal}
  {Physica A}\ }\textbf {\bibinfo {volume} {390}},\ \bibinfo {pages} {177}
  (\bibinfo {year} {2011})}\BibitemShut {NoStop}%
\bibitem [{\citenamefont {Pastor-Satorras}\ and\ \citenamefont
  {Vespignani}(2001)}]{pastor2001epidemic}%
  \BibitemOpen
  \bibfield  {author} {\bibinfo {author} {\bibfnamefont {R.}~\bibnamefont
  {Pastor-Satorras}}\ and\ \bibinfo {author} {\bibfnamefont {A.}~\bibnamefont
  {Vespignani}},\ }\href@noop {} {\bibfield  {journal} {\bibinfo  {journal}
  {Phys. Rev. Lett.}\ }\textbf {\bibinfo {volume} {86}},\ \bibinfo {pages}
  {3200} (\bibinfo {year} {2001})}\BibitemShut {NoStop}%
\bibitem [{\citenamefont {Cohen}\ \emph {et~al.}(2003)\citenamefont {Cohen},
  \citenamefont {Havlin},\ and\ \citenamefont {ben
  Avraham~Daniel}}]{cohen2003structural}%
  \BibitemOpen
  \bibfield  {author} {\bibinfo {author} {\bibfnamefont {R.}~\bibnamefont
  {Cohen}}, \bibinfo {author} {\bibfnamefont {S.}~\bibnamefont {Havlin}}, \
  and\ \bibinfo {author} {\bibnamefont {ben Avraham~Daniel}},\ }in\ \href@noop
  {} {\emph {\bibinfo {booktitle} {Handbook of Graphs and Networks}}},\
  \bibinfo {editor} {edited by\ \bibinfo {editor} {\bibfnamefont
  {S.}~\bibnamefont {Bornholdt}}\ and\ \bibinfo {editor} {\bibfnamefont
  {H.~G.}\ \bibnamefont {Schuster}}}\ (\bibinfo  {publisher} {Wiley-VCH GmbH \&
  Co.},\ \bibinfo {address} {Weinheim},\ \bibinfo {year} {2003})\ p.~\bibinfo
  {pages} {85}\BibitemShut {NoStop}%
\bibitem [{\citenamefont {Berezinskii}(1971)}]{berezinskii1971destruction}%
  \BibitemOpen
  \bibfield  {author} {\bibinfo {author} {\bibfnamefont {V.~L.}\ \bibnamefont
  {Berezinskii}},\ }\href@noop {} {\bibfield  {journal} {\bibinfo  {journal}
  {Sov. Phys. JETP}\ }\textbf {\bibinfo {volume} {32}},\ \bibinfo {pages} {493}
  (\bibinfo {year} {1971})}\BibitemShut {NoStop}%
\bibitem [{\citenamefont {Kosterlitz}\ and\ \citenamefont
  {Thouless}(1973)}]{kosterlitz1973ordering}%
  \BibitemOpen
  \bibfield  {author} {\bibinfo {author} {\bibfnamefont {J.~M.}\ \bibnamefont
  {Kosterlitz}}\ and\ \bibinfo {author} {\bibfnamefont {D.~J.}\ \bibnamefont
  {Thouless}},\ }\href@noop {} {\bibfield  {journal} {\bibinfo  {journal} {J.
  Phys. C}\ }\textbf {\bibinfo {volume} {6}},\ \bibinfo {pages} {1181}
  (\bibinfo {year} {1973})}\BibitemShut {NoStop}%
\bibitem [{\citenamefont {Costin}\ \emph {et~al.}(1990)\citenamefont {Costin},
  \citenamefont {Costin},\ and\ \citenamefont
  {Gr{\"u}nfeld}}]{costin1990infinite}%
  \BibitemOpen
  \bibfield  {author} {\bibinfo {author} {\bibfnamefont {O.}~\bibnamefont
  {Costin}}, \bibinfo {author} {\bibfnamefont {R.~D.}\ \bibnamefont {Costin}},
  \ and\ \bibinfo {author} {\bibfnamefont {C.~P.}\ \bibnamefont
  {Gr{\"u}nfeld}},\ }\href@noop {} {\bibfield  {journal} {\bibinfo  {journal}
  {J. Stat. Phys.}\ }\textbf {\bibinfo {volume} {59}},\ \bibinfo {pages} {1531}
  (\bibinfo {year} {1990})}\BibitemShut {NoStop}%
\bibitem [{\citenamefont {Bundaru}\ and\ \citenamefont
  {Gr{\"u}nfeld}(1999)}]{bundaru1999phase}%
  \BibitemOpen
  \bibfield  {author} {\bibinfo {author} {\bibfnamefont {M.}~\bibnamefont
  {Bundaru}}\ and\ \bibinfo {author} {\bibfnamefont {C.~P.}\ \bibnamefont
  {Gr{\"u}nfeld}},\ }\href@noop {} {\bibfield  {journal} {\bibinfo  {journal}
  {J. Phys. A}\ }\textbf {\bibinfo {volume} {32}},\ \bibinfo {pages} {875}
  (\bibinfo {year} {1999})}\BibitemShut {NoStop}%
\bibitem [{\citenamefont {Callaway}\ \emph {et~al.}(2001)\citenamefont
  {Callaway}, \citenamefont {Hopcroft}, \citenamefont {Kleinberg},
  \citenamefont {Newman},\ and\ \citenamefont
  {Strogatz}}]{callaway2001randomly}%
  \BibitemOpen
  \bibfield  {author} {\bibinfo {author} {\bibfnamefont {D.~S.}\ \bibnamefont
  {Callaway}}, \bibinfo {author} {\bibfnamefont {J.~E.}\ \bibnamefont
  {Hopcroft}}, \bibinfo {author} {\bibfnamefont {J.~M.}\ \bibnamefont
  {Kleinberg}}, \bibinfo {author} {\bibfnamefont {M.~E.~J.}\ \bibnamefont
  {Newman}}, \ and\ \bibinfo {author} {\bibfnamefont {S.~H.}\ \bibnamefont
  {Strogatz}},\ }\href@noop {} {\bibfield  {journal} {\bibinfo  {journal}
  {Phys. Rev. E}\ }\textbf {\bibinfo {volume} {64}},\ \bibinfo {pages} {041902}
  (\bibinfo {year} {2001})}\BibitemShut {NoStop}%
\bibitem [{\citenamefont {Dorogovtsev}\ \emph {et~al.}(2001)\citenamefont
  {Dorogovtsev}, \citenamefont {Mendes},\ and\ \citenamefont
  {Samukhin}}]{dorogovtsev2001anomalous}%
  \BibitemOpen
  \bibfield  {author} {\bibinfo {author} {\bibfnamefont {S.~N.}\ \bibnamefont
  {Dorogovtsev}}, \bibinfo {author} {\bibfnamefont {J.~F.~F.}\ \bibnamefont
  {Mendes}}, \ and\ \bibinfo {author} {\bibfnamefont {A.~N.}\ \bibnamefont
  {Samukhin}},\ }\href@noop {} {\bibfield  {journal} {\bibinfo  {journal}
  {Phys. Rev. E}\ }\textbf {\bibinfo {volume} {64}},\ \bibinfo {pages} {066110}
  (\bibinfo {year} {2001})}\BibitemShut {NoStop}%
\bibitem [{\citenamefont {Kim}\ \emph {et~al.}(2002)\citenamefont {Kim},
  \citenamefont {Krapivsky}, \citenamefont {Kahng},\ and\ \citenamefont
  {Redner}}]{kim2002infinite}%
  \BibitemOpen
  \bibfield  {author} {\bibinfo {author} {\bibfnamefont {J.-H.}\ \bibnamefont
  {Kim}}, \bibinfo {author} {\bibfnamefont {P.~L.}\ \bibnamefont {Krapivsky}},
  \bibinfo {author} {\bibfnamefont {B.}~\bibnamefont {Kahng}}, \ and\ \bibinfo
  {author} {\bibfnamefont {S.}~\bibnamefont {Redner}},\ }\href@noop {}
  {\bibfield  {journal} {\bibinfo  {journal} {Phys. Rev. E}\ }\textbf {\bibinfo
  {volume} {66}},\ \bibinfo {pages} {055101} (\bibinfo {year}
  {2002})}\BibitemShut {NoStop}%
\bibitem [{\citenamefont {Dorogovtsev}\ and\ \citenamefont
  {Mendes}(2003)}]{dorogovtsev2003evolution}%
  \BibitemOpen
  \bibfield  {author} {\bibinfo {author} {\bibfnamefont {S.~N.}\ \bibnamefont
  {Dorogovtsev}}\ and\ \bibinfo {author} {\bibfnamefont {J.~F.~F.}\
  \bibnamefont {Mendes}},\ }\href@noop {} {\emph {\bibinfo {title} {{Evolution
  of Networks: From Biological Nets to the Internet and WWW}}}}\ (\bibinfo
  {publisher} {Oxford University Press},\ \bibinfo {address} {Oxford},\
  \bibinfo {year} {2003})\BibitemShut {NoStop}%
\bibitem [{\citenamefont {Bauer}\ \emph {et~al.}(2005)\citenamefont {Bauer},
  \citenamefont {Coulomb},\ and\ \citenamefont {Dorogovtsev}}]{bauer2005phase}%
  \BibitemOpen
  \bibfield  {author} {\bibinfo {author} {\bibfnamefont {M.}~\bibnamefont
  {Bauer}}, \bibinfo {author} {\bibfnamefont {S.}~\bibnamefont {Coulomb}}, \
  and\ \bibinfo {author} {\bibfnamefont {S.~N.}\ \bibnamefont {Dorogovtsev}},\
  }\href@noop {} {\bibfield  {journal} {\bibinfo  {journal} {Phys. Rev. Lett.}\
  }\textbf {\bibinfo {volume} {94}},\ \bibinfo {pages} {200602} (\bibinfo
  {year} {2005})}\BibitemShut {NoStop}%
\bibitem [{\citenamefont {Khajeh}\ \emph {et~al.}(2007)\citenamefont {Khajeh},
  \citenamefont {Dorogovtsev},\ and\ \citenamefont
  {Mendes}}]{khajeh2007berezinskii}%
  \BibitemOpen
  \bibfield  {author} {\bibinfo {author} {\bibfnamefont {E.}~\bibnamefont
  {Khajeh}}, \bibinfo {author} {\bibfnamefont {S.~N.}\ \bibnamefont
  {Dorogovtsev}}, \ and\ \bibinfo {author} {\bibfnamefont {J.~F.~F.}\
  \bibnamefont {Mendes}},\ }\href@noop {} {\bibfield  {journal} {\bibinfo
  {journal} {Phys. Rev. E}\ }\textbf {\bibinfo {volume} {75}},\ \bibinfo
  {pages} {041112} (\bibinfo {year} {2007})}\BibitemShut {NoStop}%
\bibitem [{\citenamefont {Singh}\ \emph {et~al.}(2014)\citenamefont {Singh},
  \citenamefont {Brunson},\ and\ \citenamefont
  {Boettcher}}]{singh2014explosive}%
  \BibitemOpen
  \bibfield  {author} {\bibinfo {author} {\bibfnamefont {V.}~\bibnamefont
  {Singh}}, \bibinfo {author} {\bibfnamefont {C.~T.}\ \bibnamefont {Brunson}},
  \ and\ \bibinfo {author} {\bibfnamefont {S.}~\bibnamefont {Boettcher}},\
  }\href@noop {} {\bibfield  {journal} {\bibinfo  {journal} {Phys. Rev. E}\
  }\textbf {\bibinfo {volume} {90}},\ \bibinfo {pages} {052119} (\bibinfo
  {year} {2014})}\BibitemShut {NoStop}%
\bibitem [{\citenamefont {Nogawa}\ and\ \citenamefont
  {Hasegawa}(2014)}]{nogawa2014transition}%
  \BibitemOpen
  \bibfield  {author} {\bibinfo {author} {\bibfnamefont {T.}~\bibnamefont
  {Nogawa}}\ and\ \bibinfo {author} {\bibfnamefont {T.}~\bibnamefont
  {Hasegawa}},\ }\href@noop {} {\bibfield  {journal} {\bibinfo  {journal}
  {Phys. Rev. E}\ }\textbf {\bibinfo {volume} {89}},\ \bibinfo {pages} {042803}
  (\bibinfo {year} {2014})}\BibitemShut {NoStop}%
\bibitem [{\citenamefont {Hasegawa}\ \emph {et~al.}(2014)\citenamefont
  {Hasegawa}, \citenamefont {Nogawa},\ and\ \citenamefont
  {Nemoto}}]{hasegawa2014critical}%
  \BibitemOpen
  \bibfield  {author} {\bibinfo {author} {\bibfnamefont {T.}~\bibnamefont
  {Hasegawa}}, \bibinfo {author} {\bibfnamefont {T.}~\bibnamefont {Nogawa}}, \
  and\ \bibinfo {author} {\bibfnamefont {K.}~\bibnamefont {Nemoto}},\
  }\href@noop {} {\bibfield  {journal} {\bibinfo  {journal} {Discontinuity,
  Nonlinearity, and Complexity}\ }\textbf {\bibinfo {volume} {3}},\ \bibinfo
  {pages} {319} (\bibinfo {year} {2014})}\BibitemShut {NoStop}%
\end{thebibliography}%

\end{document}